\documentclass[structabstract]{aa}  
\usepackage{graphicx}
\usepackage{txfonts}
\usepackage{subfigure}
\begin{document}

\title{Simulating star formation in molecular cloud cores}
\subtitle{IV. The role of turbulence and thermodynamics}

\author{R. E. Attwood\inst{1}\thanks{Send offprint requests to R. E. Attwood}, S. P. Goodwin\inst{2}, D. Stamatellos\inst{1} \and A. P. Whitworth\inst{1}}

\institute{School of Physics \& Astronomy, Cardiff University, Queens Buildings, 5 The Parade, Cardiff CF24 3AA, Wales, UK.\\
\email{Rhianne.Attwood@astro.cf.ac.uk}\\
\email{Dimitrios.Stamatellos@astro.cf.ac.uk}\\
\email{Anthony.Whitworth@astro.cf.ac.uk}
\and
Department of Physics \& Astronomy, University of Sheffield, Hicks Building, Housfield Road, Sheffield S3 7RH, England, UK.\\
\email{S.Goodwin@Sheffield.ac.uk}}

\date{Received ; accepted }

\abstract
{Observations suggest that low-mass stars condense out of dense, relatively isolated, molecular cloud cores, with each core spawning a small-$N$ cluster of stars.}
{Our aim is to identify the physical processes shaping the collapse and fragmentation of a $5.4\,{\rm M}_\odot$ core, and to understand how these processes influence the mass distribution, kinematics, and binary statistics of the resulting stars.}
{We perform SPH simulations of the collapse and fragmentation of cores having different initial levels of turbulence ($\alpha_{_{\rm TURB}}=0.05,\,0.10,\,0.25$). We use a new treatment of the energy equation that captures (i) excitation of the rotational and vibrational degrees of freedom of H$_2$, dissociation of H$_2$, ionisation of H and He, and (ii) the transport of cooling radiation against opacity due to both dust and gas (including the effects of dust sublimation, molecules, and H$^-$ ions). We also perform comparison simulations using a standard barotropic equation of state.}
{We find that -- when compared with the barotropic equation of state
  -- our more realistic treatment of the energy equation results in
  more protostellar objects being formed, and a higher proportion of
  brown dwarfs; the multiplicity frequency is essentially unchanged,
  but the multiple systems tend to have shorter periods (by a factor
  $\sim 3$), higher eccentricities, and higher mass ratios. The reason for this is that small fragments are able to cool more effectively with the new treatment, as compared with the barotropic equation of state. We also note that in our simulations the process of fragmentation is often bimodal, in the following sense. The first protostar to form is usually, at the end, the most massive, i.e. the primary. However, frequently a disc-like structure subsequently forms round this primary, and then, once it has accumulated sufficient mass, quickly fragments to produce several secondaries.}
{We believe that this delayed fragmentation of a disc-like structure is likely to be an important source of very low-mass stars in nature (both low-mass hydrogen-burning stars and brown dwarf stars); hence it may be fundamental to understanding the way in which the statistical properties of stars change -- continuously but monotonically -- with decreasing mass. However, in our simulations the individual cores probably produce too many stars, and hence too many single stars. We list the physical and numerical features that still need to be included in our simulations to make them more realistic; in particular, radiative and mechanical feedback, non-ideal magneto-hydrodynamic effects, and a more sophisticated implementation of sink particles.}


\maketitle

\section{Introduction}

{\it The influence of thermodynamic effects on low-mass star formation
  in isolated, low-turbulence cores.} There is direct and indirect
observational evidence to suggest that a significant fraction of
low-mass stars form in small, relatively isolated, low-turbulence
prestellar cores, with each core spawning only a few stars. This paper
is concerned with exploring this mode of star formation, by means of
numerical simulations, with a view to evaluating (a) how the
statistical properties of the resulting protostars depend on the
initial conditions, and (b) what role is played by thermal, chemical,
and radiative processes. We adopt an analytic approach. That is
  to say, we do not seek to implement all the deterministic physical
  effects at once, and hence we do not expect at this stage to
  reproduce all the observed features of real star formation.
  Rather, we seek to establish whether particular physical effects,
  namely the thermal and radiative processes, influence the outcome in
  a systematic way. Given the complexity of star formation and the limitations of numerical codes, this incremental approach seems a more fruitful one to pursue.

{\it The direct observational evidence for low-mass star formation in isolated, low-turbulence cores.} The direct evidence for this mode of star formation comes from a number of studies. First, Andr\'e et al. (2007) have estimated the inter-core velocity dispersion in Ophiuchus (i.e. the dispersion in the bulk velocities of cores relative to their neighbours). Andr\'e et al. infer that the frequency of interactions between cores is so low that a typical core is likely to have collapsed and fragmented, internally, before it interacts with a neighbouring core. Second, estimates of the level of turbulence in low-mass prestellar cores give subsonic values, i.e. $\sigma_{_{\rm TURB}} < \sigma_{_{\rm THERM}}$ (e.g. Myers 1983; Myers et al. 1991; Myers 1998; Andr\'e et al. 2007). Indeed, Myers (1998) concludes that the decay of turbulence to subsonic levels may be a pre-requisite for the formation of low-mass protostars.

{\it The indirect observational evidence for low-mass star formation in isolated, low-turbulence cores.} Indirect evidence for this mode of star formation comes from the binary statistics of young low-mass stars, which show that a high fraction are born in binary or higher-order multiple systems. The binary fraction decreases with decreasing primary mass, and with increasing age, but for a $1\,{\rm M}_\odot$ primary it is still $\sim 60\%$ in the field (Duquennoy \& Mayor 1991). Goodwin \& Kroupa (2005) and Hubber \& Whitworth (2005) have shown that this high multiplicity requires newly-born stars to complete their early dynamical evolution in small sub-clusters containing just a few stars (i.e. ${\cal N}_{_{\rm SUBCLUSTER}}\sim 4\pm 1\;{\rm stars}$). This is because, in a small sub-cluster, $N$-body interactions tend rather quickly to deliver a tight binary, usually comprising the two most massive stars, and to eject most of the remaining stars as singles (McDonald \& Clarke 1993; Goodwin \& Kroupa 2005; Hubber \& Whitworth 2005). Therefore, if ${\cal N}_{_{\rm SUBCLUSTER}}$ is increased, a higher proportion of stars are ejected as singles, and therefore the primordial binary fraction is reduced, in contradiction with the observations.

{\it The need for large ensembles of numerical simulations.} One advantage of this mode of star formation, involving a low-mass core fragmenting in relative isolation to form a small number of protostars, is that it can be simulated with quite high numerical resolution. A related advantage is that many realisations can be computed. Given the chaotic nature of turbulent, self-gravitating gas dynamics, this is essential if robust statistical inferences are to be made.

{\it The effect of the initial level of turbulence in previous simulations.} The first comprehensive investigation of this mode of star formation was made by Goodwin et al. (2004a,b), who simulated the evolution of a large ensemble of cores, all having the same mass, initial density profile and turbulent power spectrum. They considered three different levels of turbulence, characterized by
\begin{eqnarray}\label{EQN:TURB}
\alpha_{_{\rm TURB}}&\equiv&\frac{E_{_{\rm TURB}}}{|E_{_{\rm GRAV}}|}\;\,=\;\,0.05\,,\;0.10\,,\;0.25\,,
\end{eqnarray}
where $E_{_{\rm TURB}}$ and $E_{_{\rm GRAV}}$ are the initial turbulent and gravitational energies of the core. For each value of $\alpha_{_{\rm TURB}}$, they performed many different simulations, each with a different realisation of the turbulent velocity field, in order to obtain reasonable statistics. They found that increasing the level of turbulence increased both the total number of stars formed, and the proportion of brown dwarfs.

{\it The effect of the turbulent power spectrum in previous simulations.} In a subsequent paper, Goodwin et al. (2006) considered different turbulent power spectra, of the form $P_k\propto k^{-x}$, and showed that increasing the exponent $x$ resulted in more fragments. In effect, increased $x$ means that there is more turbulent power on long wavelengths, and hence more large-scale fragmentation resulting in separate protostars. This is in accordance with the findings of Klessen \& Burkert (2001), who note that turbulence has two effects. Turbulence on very short (sub-Jeans) length-scales can be viewed as a source of extra pressure, supplementing the thermal pressure. Conversely, turbulence on larger length-scales creates the structures which, if they become sufficiently massive and dense, are amplified by self-gravity to become prestellar cores.

{\it The limitations of using a barotropic equation of state.} However, in their simulations Goodwin et al. (2004a,b, 2006) use a simple barotropic equation of state, which is designed to mimic the gross thermal behaviour of the gas at the centre of a collapsing, non-rotating $1\,{\rm M}_\odot$ protostar, as determined by Masunaga \& Inutsuka (2000). This is not realistic. First, it does not take proper account of the thermal history of protostellar gas, which depends on environment, metallicity, mass and geometry. Second, a barotropic equation of state is unable to capture thermal inertia effects, for example, the complex system of interacting pressure waves which dominates the dynamics when the gas becomes adiabatic and the thermal timescale suddenly becomes longer than the dynamical timescale. This is when fragmentation occurs (e.g. Boss et al. 2000), and thermal inertia effects are therefore critical.

{\it Our new treatment of the energy equation.} We have therefore revisited the study of Goodwin et al. (2004a,b), but now using a significantly improved treatment of the energy equation due to Stamatellos et al. (2007a). This new treatment of the energy equation captures the critical thermal and radiative effects in protostellar gas much more realistically than a barotropic equation of state. For full details of the new treatment, and of the tests that have been performed to establish its fidelity, the reader is refered to Stamatellos et al. (2007). With the new treatment of the energy equation, we find 

\begin{itemize}

\item{that the efficiency of fragmentation is increased, in the sense that on average more stars are formed, and in particular more brown dwarfs;}

\item{that the mass distribution becomes bimodal, with a peak around $0.3\;{\rm to}\;1.0\,{\rm M}_\odot$, and a subsidiary peak in the brown dwarf regime around $0.03\;{\rm to}\;0.06\,{\rm M}_\odot$ (but note that we are only considering the fragmentation products of a single core mass -- we do not expect this bimodality to translate to the overall stellar mass function, when a distribution of core masses is considered);}

\item{that fragmentation frequently involves the formation of a primary star (from material having low angular momentum) and the subsequent accumulation of a massive disc around this primary, with the disc then fragmenting ${\sim\!20,000}\;{\rm to}\;{\sim\!100,000}\;{\rm years}$ later, to produce secondaries;}

\item{that brown dwarfs result from disc fragments which {\it either} are born in the {\it outer} disc where there is not much material to assimilate, {\it or} are quickly ejected from the disc, usually by mutual interactions, before they can assimilate much mass; disc fragments that form and remain in the {\it inner} disc usually assimilate sufficient mass to become hydrogen-burning stars;}

\item{that a significant fraction of the stars end up in binary systems, or higher multiples, and these systems tend to have large mass ratios (i.e. $q\equiv M_{_2}/M_{_1}\sim 1$);}

\item{that there is little evidence for competitive accretion -- in fact, as noted above, there is a rather egalitarian process at work tending to produce binary systems with components of comparable mass.}

\end{itemize}

{\it The plan of the paper.} In Section 2, we describe the initial conditions, the constitutive physics, and the numerical method. In Section 3, we present the results of our simulations and the statistical distributions derived from them, and relate these distributions to the underlying physics. In Section 4 we summarise our conclusions.

\begin{figure*}
\centering$
\begin{array}{cccc}
\includegraphics[scale=0.2,angle=270]{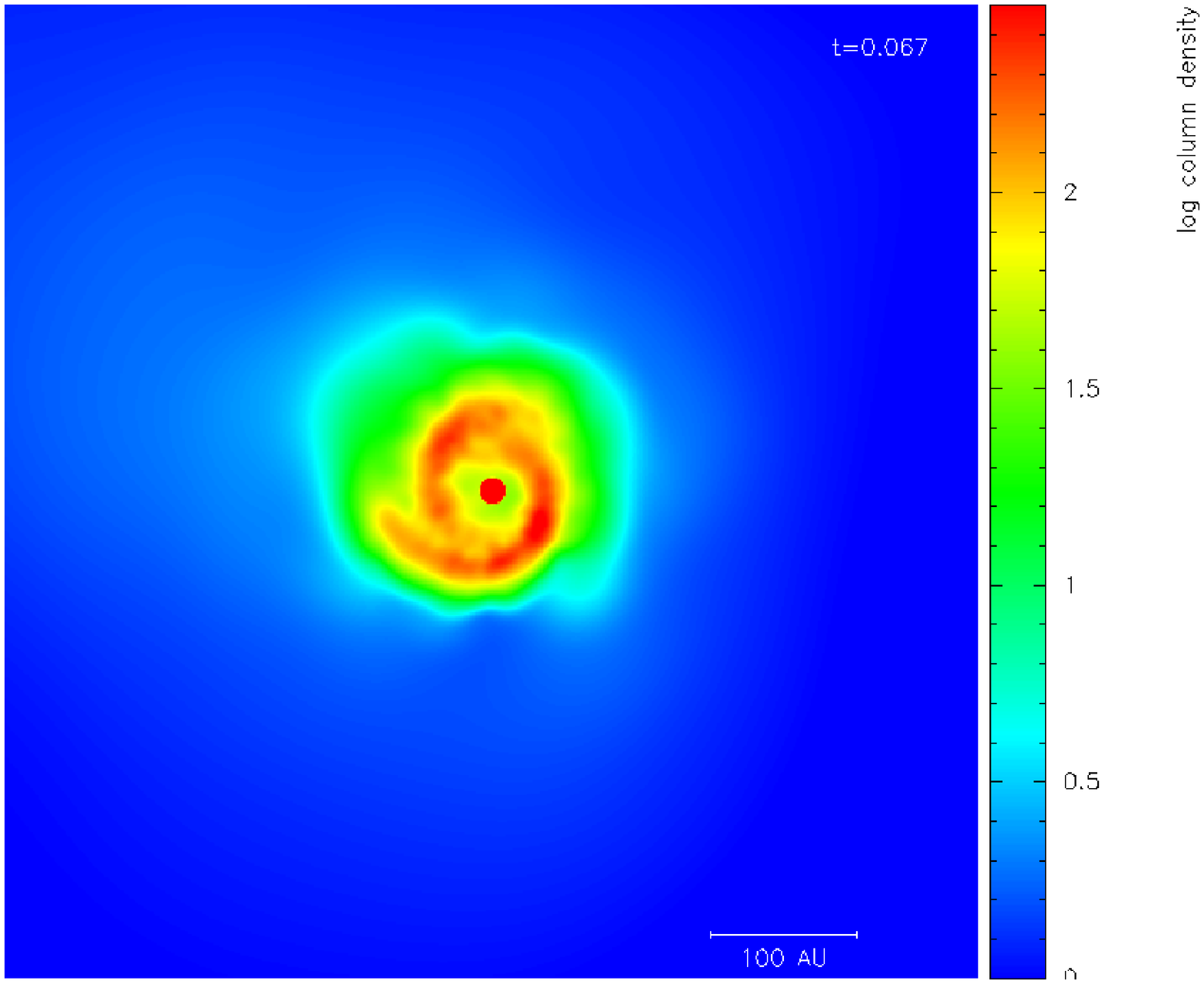} &
\includegraphics[scale=0.2,angle=270]{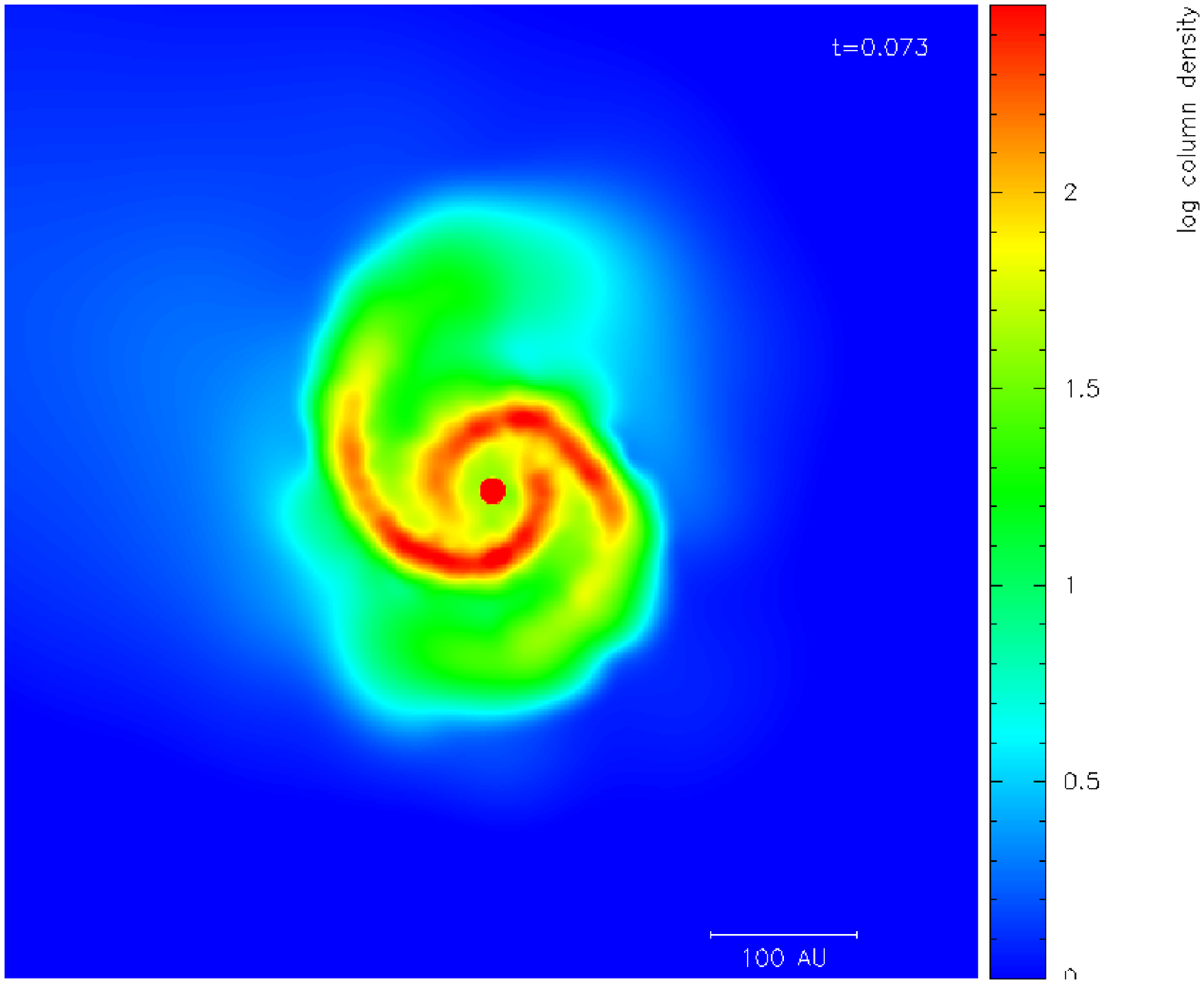} &
\includegraphics[scale=0.2,angle=270]{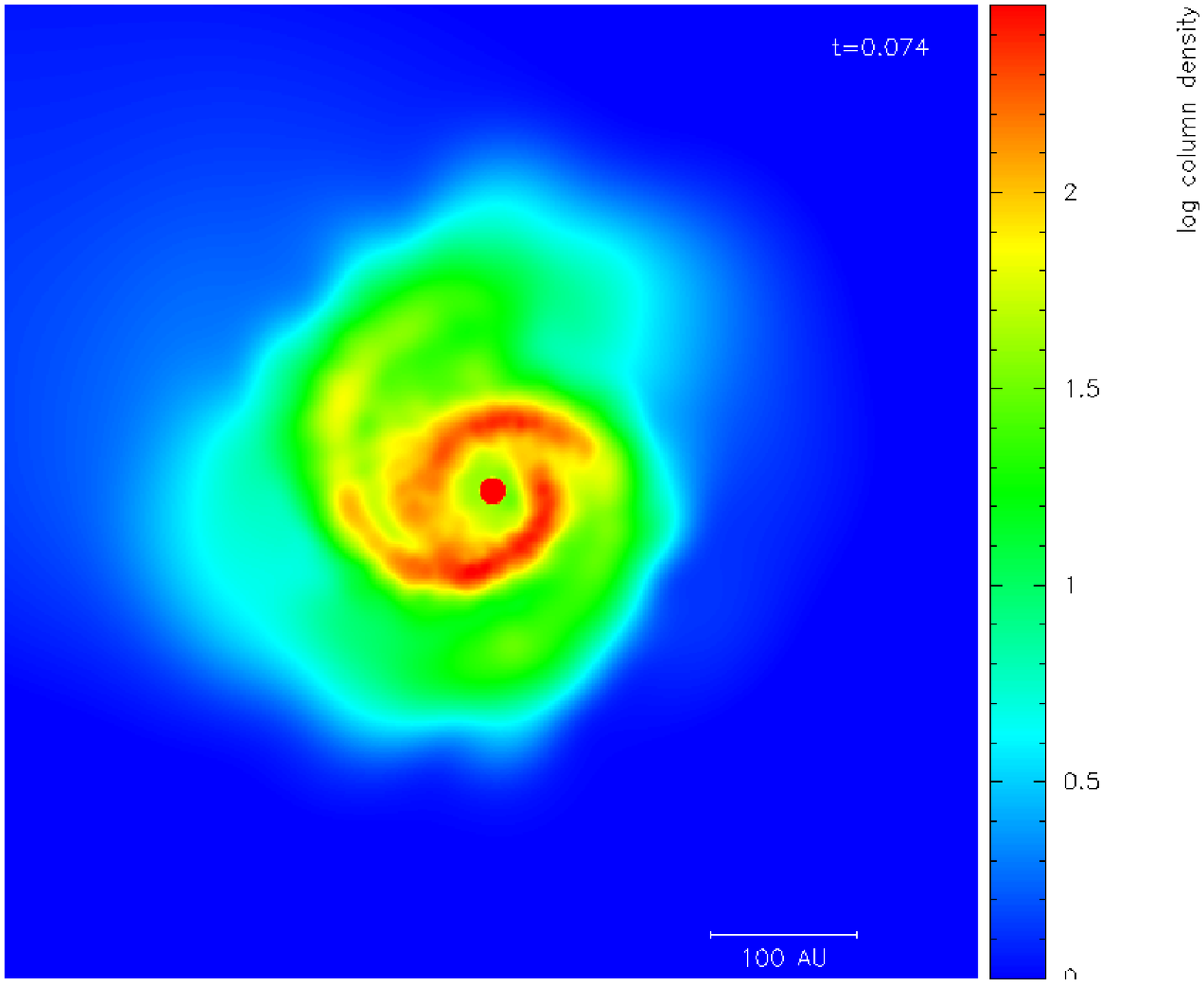} &
\includegraphics[scale=0.2,angle=270]{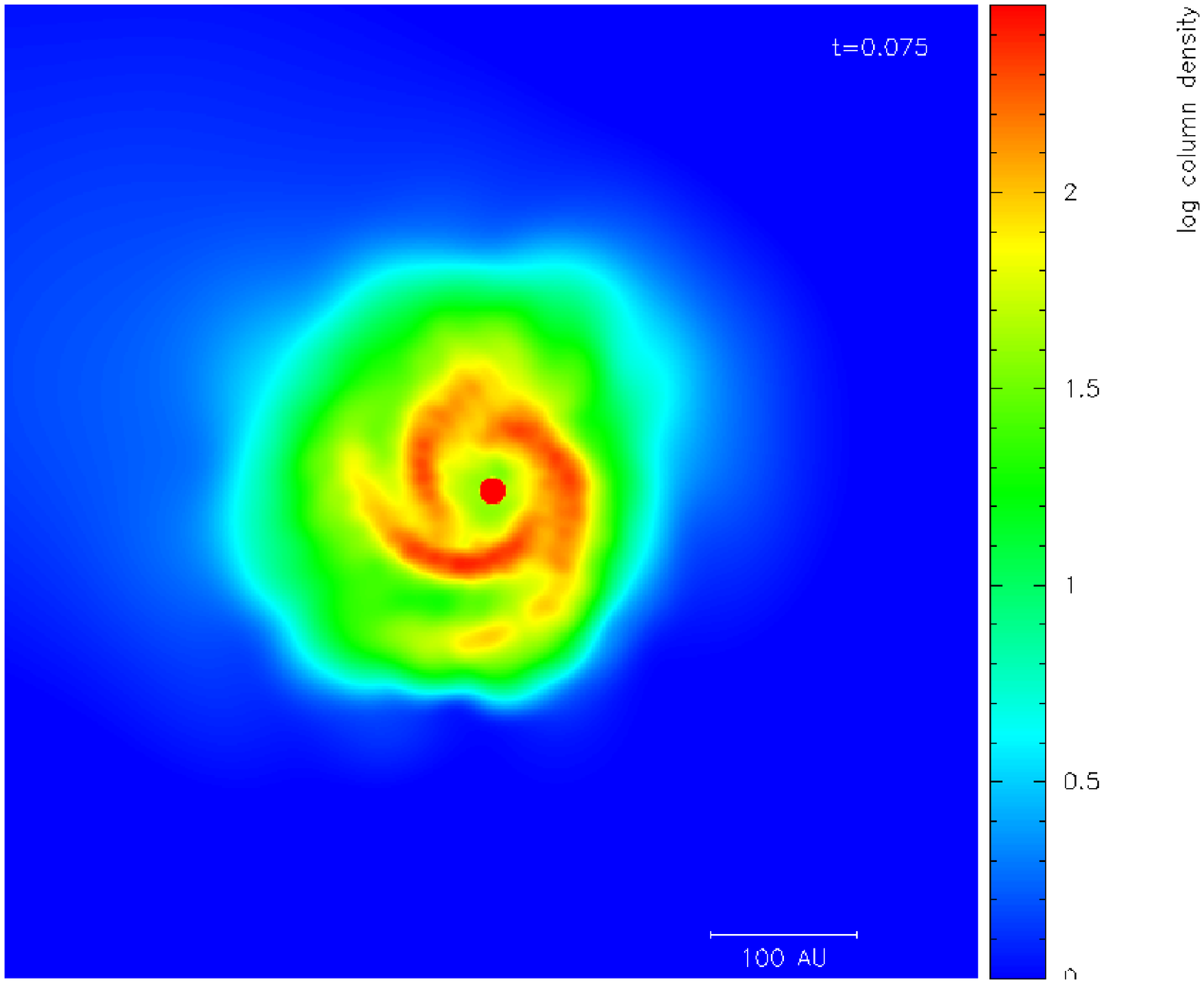} \\
\includegraphics[scale=0.2,angle=270]{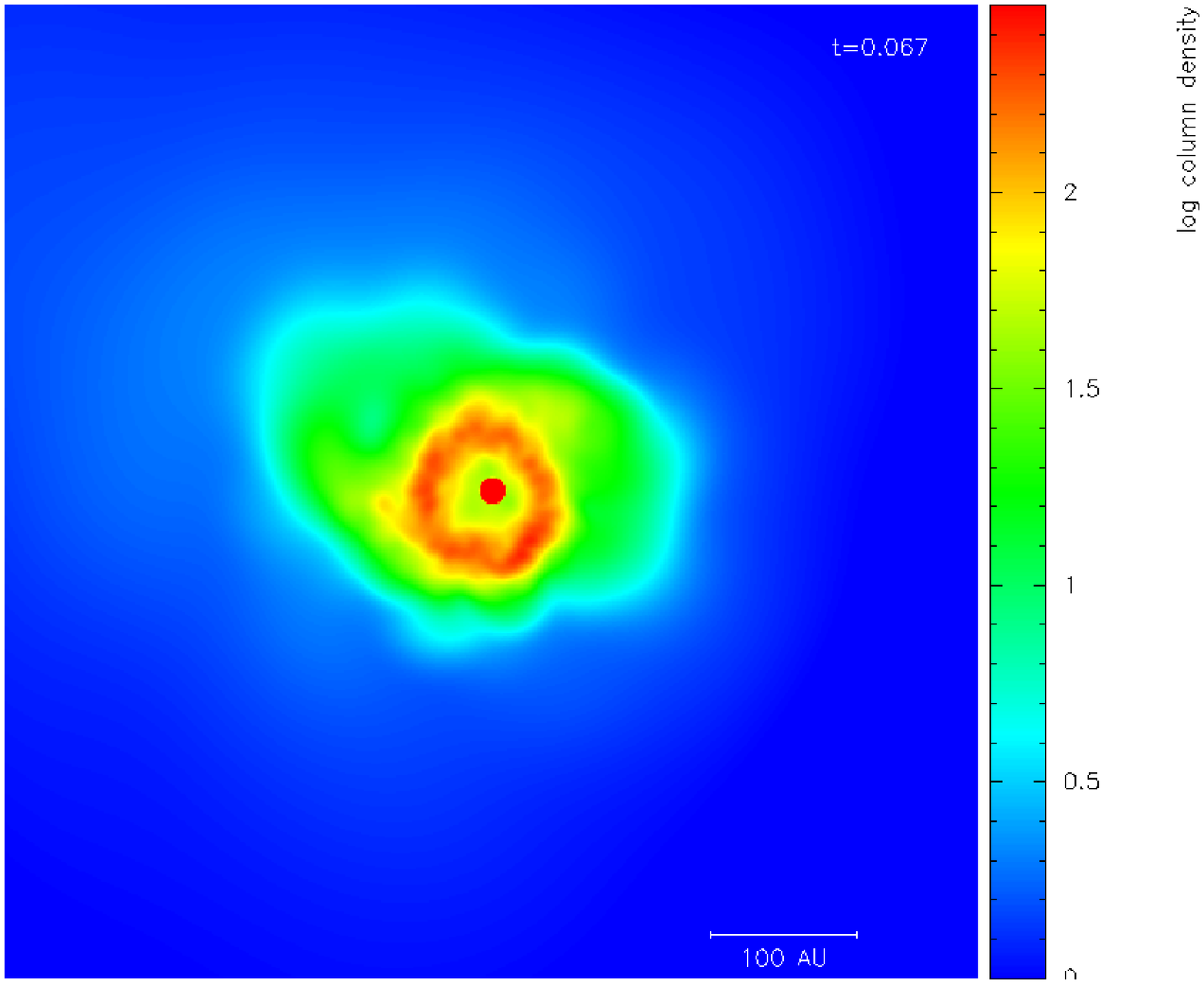} &
\includegraphics[scale=0.2,angle=270]{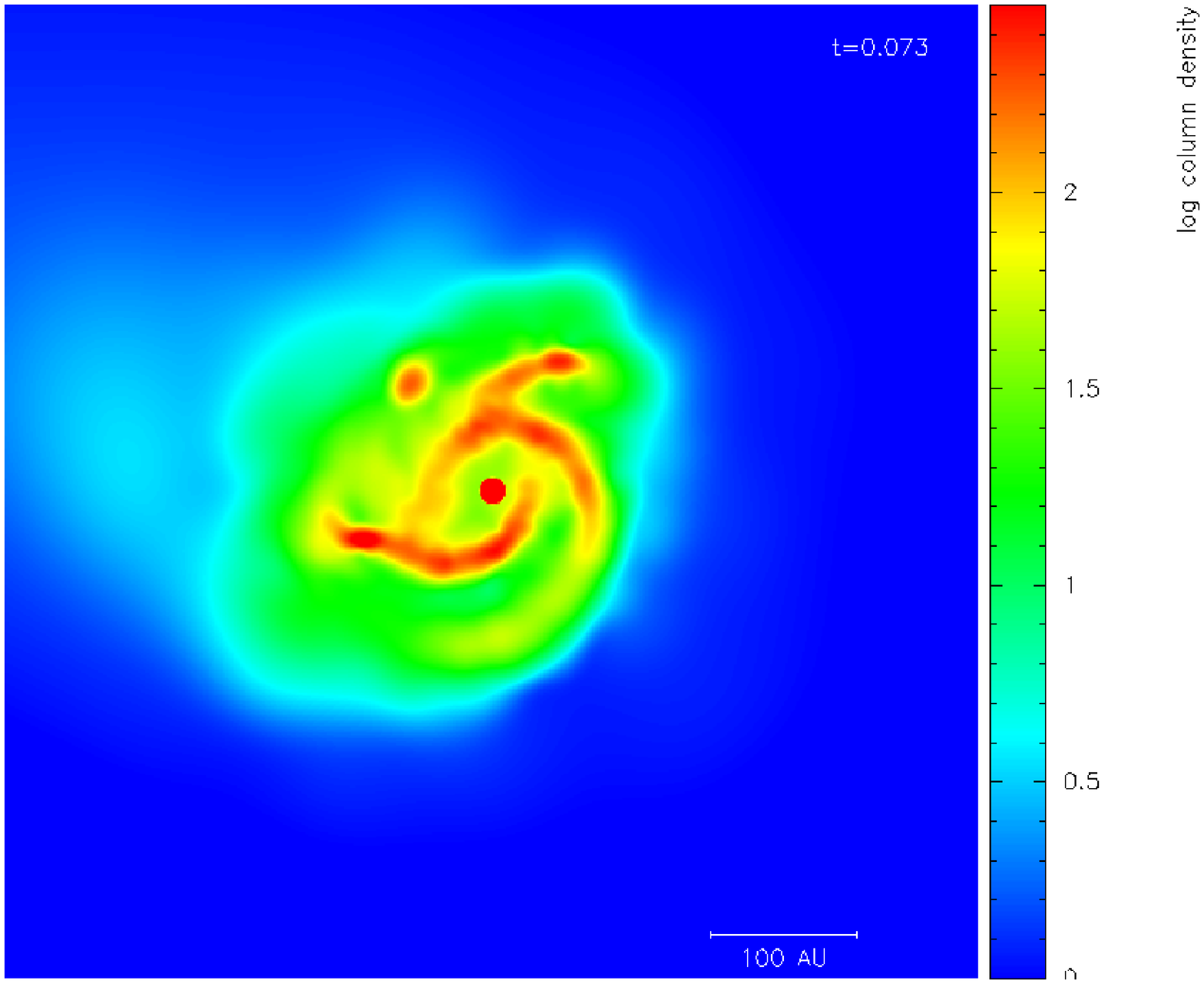} &
\includegraphics[scale=0.2,angle=270]{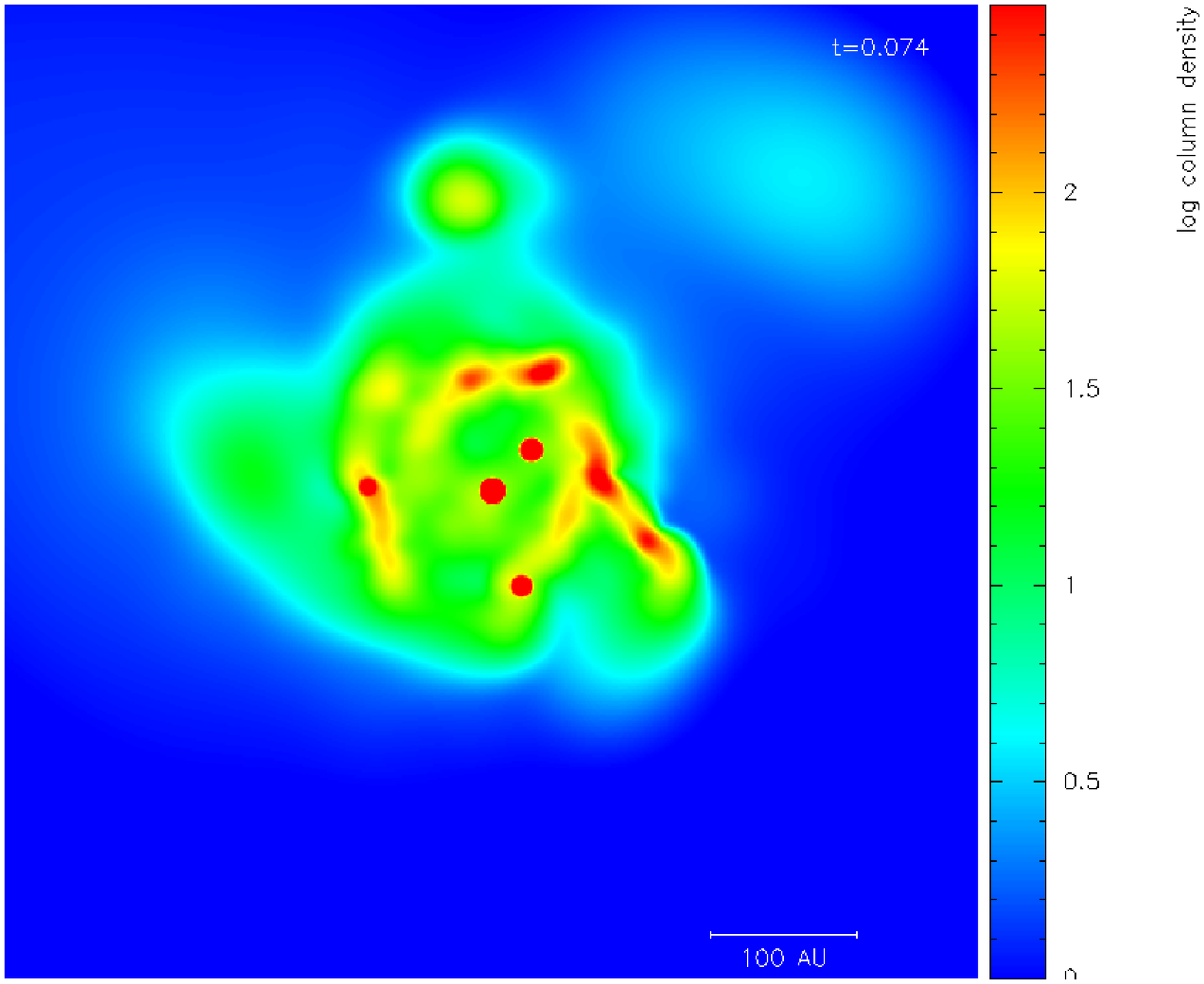} &
\includegraphics[scale=0.2,angle=270]{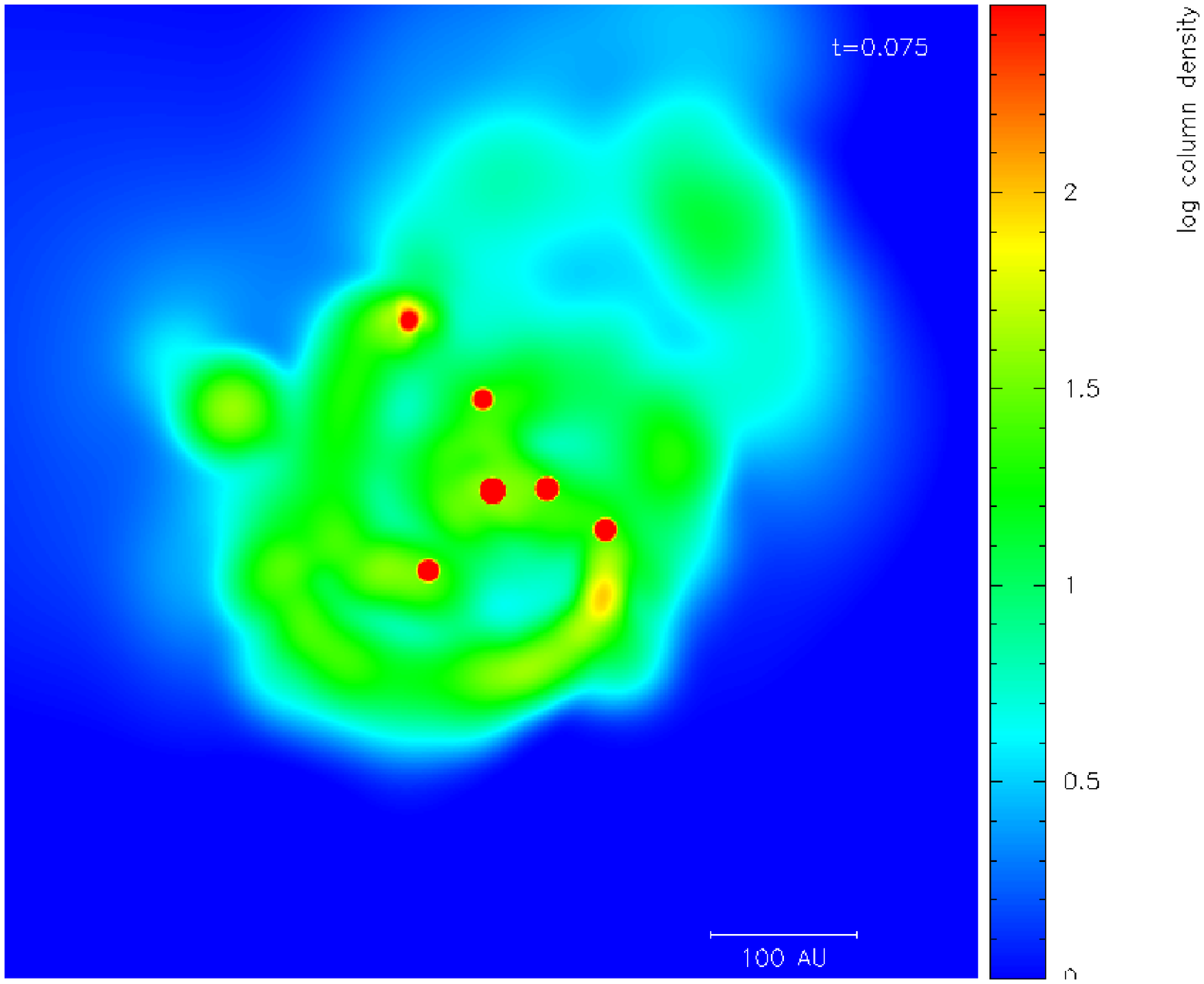} \\
\end{array}$
\caption{Two simulations of the collapse and fragmentation of a
  $5.4\,{\rm M}_\odot$ core having the same initial turbulent velocity
  field. In the first simulation (top row) the core is evolved using
  the barotropic equation of state. In the second simulation (bottom
  row) the core is evolved using the new treatment of the energy
  equation. Each frame shows the logarithm of the column density
  through the computational domain. For each simulation we show an
  image at, reading from left to right, $t=67\,{\rm kyr}\,,\;73\,{\rm
    kyr}\,,\;74\,{\rm kyr}\,,\;{\rm and}\;75\,{\rm kyr}\,$. We see
  that with the new treatment of the energy equation, the disc is more
  unstable against fragmentation.}
\label{FIG:DISCS}
\end{figure*}

\section{Simulations}

\subsection{Initial conditions}

We use the same global initial conditions as Goodwin et al. (2004a,b, 2006). These initial conditions are intended to fit the observed properties of prestellar cores like L1544 (e.g. Ward-Thompson et al. 1994; Andr{\'e} et al. 1996; Ward-Thompson et al. 1999; Andr{\'e} et al. 2000; Alves et al. 2001; Kirk et al. 2005).

{\it Density profile and mass.} The density structure of a low-mass prestellar core usually appears to consist of a central kernel having approximately uniform density, surrounded by an outer envelope in  which the density falls off radially with exponent $2 \leq |d\ell og(\rho)/d\ell og(r)| \leq 5$ . We therefore adopt a Plummer-like initial density profile,
\begin{eqnarray}
\rho(r)&=&\frac{\rho_{_{\rm KERNEL}}}{\left(1\,+\,\left(r/R_{_{\rm KERNEL}}\right)^2\right)^2}\,.
\end{eqnarray}
Here $\rho_{_{\rm KERNEL}}=3\times 10^{-18}\,{\rm g}\,{\rm cm}^{-3}$ is the central density, and $R_{_{\rm KERNEL}}=5,000\,{\rm AU}$ is the radius of the central kernel (within which the density is approximately uniform). The mass inside $R_{_{\rm KERNEL}}$ is then $M_{_{\rm KERNEL}}=1.1\,{\rm M}_\odot$. The outer envelope of the core extends out to $R_{_{\rm CORE}}=50,000\,{\rm AU}$, so the total core mass is $M_{_{\rm CORE}}=5.4\,{\rm M}_\odot$ and the density at the boundary of the core is $10^4$ times lower than at the centre.

{\it Temperature.} The gas is initially isothermal at $T=10\,{\rm K}$, and hence the initial ratio of thermal to gravitational energy is
\begin{eqnarray}
\alpha_{_{\rm THERM}}&\equiv&\frac{E_{_{\rm THERM}}}{|E_{_{\rm GRAV}}|}\;\,=\;\,0.3\,.
\end{eqnarray}

{\it Turbulence.} We also impose an initial divergence-free Gaussian random velocity field on each core. We set the power spectrum of this velocity field to be $P_kdk\propto k^{-4}dk$, since this mimics the scaling laws observed in molecular clouds (e.g. Larson 1981; Burkert \& Bodenheimer 2000). This prescription for initialising the velocity field is normally referred to as {\it turbulence} (c.f. Bate et al. 2002a,b, 2003; Bonnell et al. 2003; Goodwin et al. 2004a,b, 2006), and we shall follow this convention. However, we emphasise that this is not self-consistent, fully developed turbulence (c.f. Offner et al. 2008), it is simply a device used to seed the core with substructure, which may then be dissipated or become amplified by self-gravity. In the simulations presented here, we consider quite low levels of turbulence, characterised by $\alpha_{_{\rm     TURB}}=0.05,\,0.10,\,0.25$ (where $\alpha_{_{\rm TURB}}$ is defined in Eqn. \ref{EQN:TURB}). We note that these are typical values for the level of turbulence in observed low-mass cores, as collated in the catalogue of Jijina, Myers \& Adams (1999); they are much lower than the value $\alpha_{_{\rm TURB}}=1\;$ used by Bate et al. (2002a,b, 2003) and Bonnell et al. (2003) in their simulations of more massive cores.

\subsection{Constitutive physics}

{\it Barotropic equation of state.} For comparison purposes, we follow Goodwin et al. (2004a,b, 2006) in using a barotropic equation of state, such that the isothermal sound speed, $c_{_{\rm S}}$, is given by
\begin{eqnarray}\label{EQN:BAROTROPIC}
c_{_{\rm S}}^2(\rho)&\equiv&\frac{P(\rho)}{\rho}\;=\;c_{_{\rm O}}^2\,\left\{1\,+\,\left(\frac{\rho}{\rho_{_{\rm O}}}\right)^{2/3}\right\}\,.
\end{eqnarray}
Here $c_{_{\rm O}}=0.2\,{\rm km}\,{\rm s}^{-1}$ and $\rho_{_{\rm O}}=10^{-13}\,{\rm g}\,{\rm cm}^{-3}$. At low densities ($\rho<\rho_{_{\rm O}}$) the gas is approximately isothermal; $c_{_{\rm O}}=0.2\,{\rm km}\,{\rm s}^{-1}$ corresponds to a mix of $X=0.70$ molecular hydrogen and $Y=0.28$ atomic helium at $T=10\,{\rm K}$. At higher densities ($\rho>\rho_{_{\rm O}}$) the presumption is that the gas becomes opaque to its own cooling radiation, and heats up adiabatically. Until the temperature rises above $\sim 200\,{\rm K}$, the rotational degrees of freedom of molecular hydrogen are not significantly excited, so the effective adiabatic exponent is $\gamma\sim 5/3$ and the sound speed rises as $\sim\!\rho^{1/3}$. This in turn means that the Jeans mass increases rather rapidly once the density exceeds $\rho_{_{\rm CRIT}}$, roughly as $M_{_{\rm JEANS}}\propto\rho^{1/2}$. We note that the simulations of Bate et al. (2002a,b, 2003) and Bonnell et al. (2003) use a similar barotropic equation of state to us, but with $\gamma=7/5$ in the adiabatic regime; consequently their Jeans mass increases much more slowly with increasing density, $M_{_{\rm JEANS}}\propto\rho^{1/10}$, giving a greatly extended window of opportunity for fragmentation at low masses.

{\it The need for a more realistic treatment of the energy equation.} Eqn. (\ref{EQN:BAROTROPIC}) is a reasonably good fit to the run of temperature with density at the centre of a collapsing, non-rotating $1\,{\rm M}_\odot$ protostar, as obtained by the detailed computations of Masunaga \& Inutsuka (2000). It also matches reasonably well the earlier results of Larson (1969) and Tohline (1982) for the thermodynamics of protostellar matter. However, even if one limits consideration to the collapse of a spherically symmetric, non-rotating $1\,{\rm M}_\odot$ protostar, the run of temperature with density away from the centre shows quite a large variance, relative to the central values. Moreover, as soon as the simulations involve condensations which have masses below $1\,{\rm M}_\odot$ and/or are non-spherical, these condensations will tend to become opaque, and to heat up, at a significantly higher density. This is because in general (a) the optical depths involved are lower, and (b) the rates of compressional heating are slower. A barotropic equation of state cannot capture these effects, nor can it handle situations where the thermal timescale becomes comparable with the dynamical one. Yet these are precisely the circumstances under which fragmentation occurs, as evidenced by the simulations of Boss et al. (2000). It is therefore of great interest to explore what happens when the energy equation and associated radiative transport are treated more realistically.

{\it New treatment of the energy equation.} Stamatellos et al. (2007a)
have introduced a new treatment of the energy equation and the
associated radiative transport, which captures all the above effects,
but incurs a very small computational overhead ($\sim 3\%$). This
method incorporates the internal energy associated with the rotational
and vibrational degrees of freedom of H$_{_2}$, the dissociation of
H$_{_2}$, and the ionisation of H$^{\rm o}$, He$^{\rm o}$ and
He$^+$. It also treats the transport of heating and cooling radiation, using an opacity which -- in different regimes -- is delivered by refractory dust cores with ice mantles, refractory dust cores without ice mantles, molecules, bound/free and free/free transitions and electron scattering. The background radiation field and the metallicity are user-defined free parameters. In the current version it is assumed that the dust abundance is proportional to the metallicity and that the dust properties do not evolve, but these assumptions can easily be relaxed. The method adapts efficiently to circumstances in which the thermal timescale is shorter than the dynamical one, and vice versa. The method is able to handle optically thin, intermediate and optically thick situations, and it has been extensively tested against detailed computations (like those of Masunaga \& Inutsuka 2000) and analytic solutions (like those of Spiegel 1957). For full details of the method, and of the tests to which it has been subjected, the reader is referred to Stamatellos et al. (2007a).

\subsection{Numerical method}

{\it The SPH code.} The simulations are performed using the {\sc dragon} smoothed particle hydrodynamics code (Goodwin et al. 2004a). This is a standard SPH code (e.g. Monaghan 1992), and has been extensively tested and optimised. It uses a second-order Runge-Kutta integration scheme, multiple-particle timesteps evaluated according to a Courant-Friedrich type condition, adaptive smoothing lengths and standard artificial viscosity (Monaghan 1992). An octal spatial decomposition tree (Barnes \& Hut 1986) is used to facilitate the collation of neighbour lists and the computation of gravitational accelerations. The smoothing lengths of the particles are adjusted so that each particle always has exactly ${\cal N}_{_{\rm NEIB}}=50$ neighbours. In this way numerical diffusion is kept very low (Attwood et al. 2007).

{\it Sink particles.} During SPH simulations of star formation, the timestep required to follow the motion of the gas decreases as the stellar condensations increase in density, and eventually all the computational resources are being used to follow the first one or two condensations to form. The rest of the simulation then ends up in a state of suspended animation, and so it is impossible to know whether any further stars would have formed, or what their properties might have been. In order to overcome this problem, we invoke sinks (Bate et al. 1995). Specifically, a sink is created if (a) the density of one of the SPH particles, $i$, exceeds the threshold $\rho_{_{\rm SINK}}=10^{-11}\,{\rm g}\,{\rm cm}^{-3}$; (b) particle $i$ has negative velocity divergence; and (c) particle $i$ and its neighbours have net negative energy (i.e. are bound). The newly-formed sink incorporates all the neighbours of particle $i$ within a distance $R_{_{\rm SINK}}=5\,{\rm AU}$. Any SPH particle which subsequently passes within a distance $R_{_{\rm SINK}}$ of a sink and is bound to that sink, is assimilated by the sink. We record and update the total mass, position, velocity and angular momentum of each sink. The sinks created in a simulation are identified as stars. During post-run analysis, we can use their masses, positions and velocities to determine the mass distribution, kinematics and binary statistics of stars.

{\it Resolution.} All the simulations reported here are performed with ${\cal N}_{_{\rm TOT}}=25,000$ SPH particles. Hence the mass resolution is
\begin{eqnarray}
M_{_{\rm MIN}}&\sim&\frac{{\cal N}_{_{\rm NEIB}}\,M_{_{\rm TOT}}}{{\cal N}_{_{\rm TOT}}}\;\sim\;0.01\,{\rm M}_\odot\,,
\end{eqnarray}
(Bate \& Burkert 1997), and the formation of stars below this mass is inhibited (Whitworth 1998). Since sink particles with radius $R_{_{\rm SINK}}=5\,{\rm AU}$ are created above a threshold density, $\rho_{_{\rm SINK}}=10^{-11}\,{\rm g}\,{\rm cm}^{-3}$, the minimum linear resolution is $R_{_{\rm SINK}}=5\,{\rm AU}$. The discs which form in our simulations typically have radius $R_{_{\rm DISC}}\sim 50\,{\rm AU}$ and half-thickness $Z_{_{\rm DISC}}\sim 5\,{\rm AU}$, so they are not resolved in the vertical direction. However, the fragmentation of a disc is essentially a two-dimensional process, in the sense that the forces which drive the accumulation of matter into a proto-fragment (the forces which enter into the Toomre criterion) are in the plane of the disc. Therefore we do not believe that the limited resolution is critical to our conclusions (but see Nelson 2006).

\subsection{Nomenclature}

{\it Stars.} We will refer to all sinks as {\it stars}, irrespective of their mass. The implication is that all objects which form by gravitational instability, on a dynamical timescale, are ultimately stars, irrespective of whether they have sufficient mass to burn hydrogen or deuterium. We note that for gas with isothermal sound speed $c_{_{\rm S}}\sim 0.2\,{\rm km}\,{\rm s}^{-1}$ the dynamical timescale (i.e. the freefall time for a marginally Jeans unstable fragment) is rather short,
\begin{eqnarray}
t_{_{\rm DYN}}&\simeq&\frac{G\,M}{10\,c_{_{\rm S}}^3}\;\simeq\;60\,{\rm kyr}\,\left(\frac{M}{{\rm M}_\odot}\right)\,,
\end{eqnarray}
particularly for very low masses.

{\it Planets.} In contrast, objects which form by core accretion of solid matter (and possibly the subsequent acquisition of a gaseous envelope), on a much longer timescale (typically more than $1\,{\rm Myr}$), are planets. Such objects cannot form in the present simulations, because the physics of dust settling and aggregation is not included, and the duration of the simulations is too short.

{\it The Stellar Initial Mass Function.} In the context of star formation it is entirely appropriate to discuss -- as a single group -- all objects which form by gravitational instability. It is appropriate because the Initial Mass Function (IMF) appears to be continuous across the hydrogen-burning limit at $\sim 0.080\,{\rm M}_\odot$ (e.g. Chabrier 2003), and there is no evidence for its not also being continuous across the deuterium-burning limit at $\sim 0.013\,{\rm M}_\odot$ (just very poor statistics). It would therefore seem both more sensible and more practical to refer to this distribution of mass as the {\it Stellar IMF}, rather than the {\it IMF for Stars, Brown Dwarfs and Planetary-Mass Objects that Form by Gravitational Instability}. The really critical issue here is that, in the mass interval $0.001\;{\rm to}\;0.010\,{\rm M}_\odot$, there are probably both objects that should be called stars and objects that should be called planets; distinguishing them observationally could be rather hard.

{\it The continuity of statistical properties across the hydrogen-burning limit.} It is also appropriate because all the other observational evidence suggests that the statistical properties of young brown-dwarf stars and low-mass hydrogen-burning stars (i.e.  kinematics, clustering properties, binary statistics, frequency and lifetimes of circumstellar discs, accretion rates and outflow rates, etc.) form a continuum (e.g. Burgasser et al. 2007; Luhmann et al. 2007; Whitworth et al. 2007).

{\it Continuity of formation mechanisms across the hydrogen-burning limit.} This continuity of statistical properties suggests that the bulk of a star's mass is committed to the star it ends up in before it discovers which nuclear fuels it can burn and which it cannot. In other words, it seems likely that the mechanisms which form hydrogen-burning stars of mass $\sim 0.090\,{\rm M}_\odot$ also work to form brown-dwarf stars of mass $\sim 0.070\,{\rm M}_\odot$, and vice versa.

{\it Trends across the hydrogen-burning limit.} That is not to say that there are not strong trends in the statistical properties of stars as the mass decreases to very low values. Indeed, one of the main conclusions of this paper is that as one crosses the hydrogen burning limit towards lower masses, (a) the proportion of stars that are formed by disc fragmentation is steadily increasing, and (b) this is reflected in their statistical properties.

\subsection{Ensembles of simulations for different treatments of the thermodynamics and different initial levels of turbulence}

For each value of $\alpha_{_{\rm TURB}}\;(=0.05,\,0.10,\,0.25)$, we have performed an ensemble of 20 simulations using the barotropic equation of state (Eqn. \ref{EQN:BAROTROPIC}), and an ensemble of 20 simulations using our new treatment of the energy equation (Stamatellos et al. 2007). This gives a total of 120 simulations. The 20 simulations in an ensemble --- i.e. the set of 20 simulations all using the same treatment of the thermodynamics (barotropic equation of state {\it or} new treatment of the energy equation) and the same initial level of turbulence (same $\alpha_{_{\rm TURB}}$) --- are distinguished solely by having different realisations of the turbulent velocity field. This simply requires the use of a different random-number seed to generate the initial turbulent velocity field.

Each simulation is evolved for a total of $300\,{\rm kyr}$. For comparison, the initial freefall time at the centre of the core is $t_{_{\rm FF}}\simeq 40\,{\rm kyr}$; and the first star usually forms after $50\;{\rm to}\;70\,{\rm kyr}$.

\section{Results}

Table \ref{TAB:BAROTROPIC} lists, for each simulation performed with
the barotropic equation of state, the identifier ({\sc id}); the
initial level of turbulence ($\alpha_{_{\rm TURB}}$), the total mass
which ends up in stars ($\sum\{M_{\!\star}\}/{\rm M}_\odot$), the
total number of stars (${\cal N}_{\!\star}$), and the total number of
brown dwarfs (${\cal N}_{_{\rm BD}}$), at the end of the simulation
(after $300\,{\rm kyr}$); the types of multiple system that have
formed (S = single, B = binary, T = triple, Q = quadruple); and the masses of the individual stars ($M_{\!\star}/{\rm M}_\odot$; with a superscript indicating which ones are components of multiple systems). Table \ref{TAB:NEWENEQN} lists the same information for the simulations performed using the new treatment of the energy equation. To save space, we give here only the first twenty lines of each table (i.e. the results obtained with $\alpha_{_{\rm TURB}}=0.05$); the full tables are available online. 

\begin{table*}
\caption[]{Results of the simulations performed using the barotropic equation of state with $\alpha_{_{\rm TURB}}=0.05,\,0.10\;{\rm and}\;0.25$, at time $t=0.3\,{\rm Myr}$.}
\label{TAB:BAROTROPIC}
\begin{tabular}{lllllll}\hline&&&&&\\
{\sc ID} & $\alpha_{_{\rm TURB}}$ & $\sum\{M_{\!\star}\}/{\rm M}_\odot$ & ${\cal N}_{\!\star}$ &
${\cal N}_{_{\rm BD}}$ & Mult & $M_{\!\star}\!/{\rm M}_{\odot}$ \\
&&&&&\\\hline 
N071 & 0.05 & 3.731 & 4  & 0 & T    & 1.297$^t$, 0.970, 0.749$^t$, 0.715$^t$ \\
N072 & 0.05 & 3.867 & 5  & 0 & T    & 1.638$^t$, 1.204$^t$, 0.472$^t$, 0.285, 0.268 \\
N073 & 0.05 & 3.282 & 5  & 0 & B    & 1.216$^b$, 1.111$^b$, 0.386, 0.186, 0.383 \\
N074 & 0.05 & 4.001 & 7  & 3 & Q    & 1.174$^q$, 1.106$^q$, 0.811$^q$, 0.806$^q$, 0.049, 0.031, 0.024 \\
N075 & 0.05 & 3.307 & 4  & 0 & T    & 0.956$^t$, 0.911$^t$, 0.728$^t$, 0.712 \\
N076 & 0.05 & 3.915 & 7  & 2 & Q    & 1.001$^q$, 0.914$^q$, 0.694, 0.593$^q$, 0.583$^q$, 0.088, 0.042 \\
N077 & 0.05 & 3.646 & 1  & 0 & S    & 3.646 \\
N078 & 0.05 & 3.814 & 6  & 0 & T    & 0.998$^t$, 0.892$^t$, 0.854$^t$, 0.763, 0.199, 0.108 \\
N079 & 0.05 & 3.959 & 3  & 0 & T    & 2.319$^t$, 0.823$^t$, 0.817$^t$ \\
N080 & 0.05 & 3.700 & 1  & 0 & S    & 3.700 \\
N081 & 0.05 & 3.690 & 1  & 0 & S    & 3.690 \\
N082 & 0.05 & 3.928 & 5  & 2 & T    & 1.322$^t$, 1.286$^t$, 1.214$^t$, 0.070, 0.036 \\
N083 & 0.05 & 3.905 & 2  & 0 & B    & 2.459$^b$, 1.446$^b$ \\
N084 & 0.05 & 3.987 & 2  & 0 & B    & 3.056$^b$, 0.931$^b$ \\
N085 & 0.05 & 3.911 & 2  & 0 & B    & 2.151$^b$, 1.760$^b$ \\
N086 & 0.05 & 3.774 & 6  & 1 & Q    & 1.037$^q$, 1.007$^q$, 0.741$^q$, 0.732$^q$, 0.219, 0.038 \\
N087 & 0.05 & 3.404 & 1  & 0 & S    & 3.404 \\ 
N088 & 0.05 & 3.874 & 1  & 0 & S    & 3.874 \\
N089 & 0.05 & 3.491 & 5  & 0 & B    & 1.005$^b$, 0.934$^b$, 0.695, 0.693, 0.164 \\
N090 & 0.05 & 3.778 & 1  & 0 & S    & 3.778 \\
\hline 
\end{tabular}
\end{table*}

\begin{table*}
\caption[]{Results of the simulations using the new treatment of the
  energy equation with $\alpha_{_{\rm TURB}}=0.05,\,0.10\;{\rm and}\;0.25$, at time $t=0.3\,{\rm Myr}$.}
\label{TAB:NEWENEQN}
\begin{tabular}{lllllll}\hline&&&&&&\\
{\sc ID} & $\alpha_{_{\rm TURB}}$ & $\sum\{M_{\!\star}\}/{\rm M}_\odot$ & ${\cal N}_{\!\star}$ &
${\cal N}_{_{\rm BD}}$ & Mult & $M_{\!\star}\!/{\rm M}_{\odot}$ \\
&&&&&\\\hline 
T071 & 0.05 & 3.161 & 8  & 3 & B    & 0.825$^b$, 0.810$^b$, 0.622, 0.494, 0.346, 0.029, 0.020, 0.015 \\
T072 & 0.05 & 2.212 & 6  & 1 & BT   & 0.750$^b$, 0.603$^b$, 0.281$^t$, 0.279$^t$, 0.275$^t$, 0.024 \\
T073 & 0.05 & 3.200 & 5  & 1 & B    & 0.877$^b$, 0.870$^b$, 0.696, 0.678, 0.079 \\
T074 & 0.05 & 3.561 & 13 & 6 & BT   & 0.830$^b$, 0.828$^b$, 0.439$^t$, 0.432$^t$, 0.427$^t$, 0.328, 0.154, 0.039, 0.034, 0.023, 0.013, 0.007, \\
                                                                                                            & & & & & & \hspace{10.1cm} 0.007 \\
T075 & 0.05 & 3.252 & 7  & 2 & BB   & 0.753$^{b1}$, 0.748$^{b1}$, 0.599$^{b2}$, 0.553, 0.546$^{b2}$, 0.032, 0.021\\
T076 & 0.05 & 3.918 & 9  & 5 & Q    & 1.118$^q$, 1.104$^q$, 0.722$^q$, 0.715$^q$, 0.079, 0.070, 0.056, 0.029, 0.025 \\
T077 & 0.05 & 3.884 & 7  & 3 & B    & 1.338$^b$, 1.149$^b$, 0.702, 0.635, 0.032, 0.017, 0.011 \\
T078 & 0.05 & 3.559 & 12 & 3 & T    & 0.679, 0.678$^t$, 0.505$^t$, 0.478$^t$, 0.435, 0.261, 0.191, 0.168, 0.087, 0.040, 0.023, 0.014 \\
T079 & 0.05 & 3.410 & 10 & 1 & T    & 0.733$^t$, 0.722, 0.487$^t$, 0.481$^t$, 0.461, 0.186, 0.139, 0.095, 0.088, 0.018 \\
T080 & 0.05 & 3.894 & 4  & 1 & B    & 1.480$^b$, 1.237, 1.159$^b$, 0.018 \\
T081 & 0.05 & 3.613 & 5  & 1 & B    & 1.158$^b$, 1.059$^b$, 0.686, 0.645, 0.065 \\
T082 & 0.05 & 3.741 & 7  & 2 & T    & 0.907, 0.884$^t$, 0.758$^t$, 0.754$^t$, 0.399, 0.026, 0.013 \\
T083 & 0.05 & 3.739 & 5  & 1 & B    & 0.999, 0.983$^b$, 0.912$^b$, 0.815, 0.030 \\
T084 & 0.05 & 3.868 & 8  & 3 & B    & 1.219$^b$, 1.191$^b$, 1.123, 0.139, 0.092, 0.051, 0.031, 0.022 \\
T085 & 0.05 & 3.749 & 7  & 2 & B    & 0.882$^b$, 0.850, 0.832, 0.804$^b$, 0.332, 0.026, 0.023 \\
T086 & 0.05 & 3.898 & 6  & 1 & Q    & 0.863$^q$, 0.850$^q$, 0.849$^q$, 0.840$^q$, 0.460, 0.036 \\
T087 & 0.05 & 3.434 & 1  & 0 & S    & 3.434 \\ 
T088 & 0.05 & 3.381 & 8  & 3 & BB   & 1.435$^{b1}$, 1.073$^{b1}$, 0.332$^{b2}$, 0.320$^{b2}$, 0.164, 0.021, 0.019, 0.017 \\
T089 & 0.05 & 3.259 & 9  & 2 & B    & 0.645$^b$, 0.617$^b$, 0.612, 0.542, 0.342, 0.314, 0.146, 0.022, 0.019 \\
T090 & 0.05 & 3.811 & 7  & 2 & T    & 2.100$^t$, 1.032$^t$, 0.248$^t$, 0.187, 0.122, 0.058, 0.034 \\
\hline 
\end{tabular}
\end{table*}

\subsection{Efficiency and timing of star formation}

\begin{table*}
\caption{Basic statistics for each treatment of the thermodynamics (barotropic equation of state or new treatment of the energy equation) and each value of the initial level of turbulence ($\alpha_{_{\rm TURB}}$).}
\label{TAB:GLOBAL}      
\centering          
\begin{tabular}{lcccccccccccccc}\hline
& & & & & & & & & & & & & & \\
{\sc thermodynamics} & \hspace{0.35cm} & $\alpha_{_{\rm TURB}}$ & ${\cal N}_{_{\rm REAL}}$ & $\eta$ & $\overline{\cal N}_\star$ & \hspace{0.35cm} & $S$ & $B$ & $T$ & $Q$ & \hspace{0.35cm} & {\bf mf} & {\bf cp} & {\bf cf} \\
& & & & & & & & & & & & & & \\\hline
& & & & & & & & & & & & & & \\
{\sc barotropic equation of state} & & 0.05 & 20 & 0.694 & 3.45 & & 29 & 5 & 6 & 3 & & 0.33 & 0.58 & 1.19 \\
& & 0.10 & 20 & 0.658 & 4.15 & & 30 & 2 & 7 & 7 & & 0.35 & 0.64 & 1.57 \\
& & 0.25 & 20 & 0.600 & 5.05 & & 27 & 2 & 10 & 10 & & 0.46 & 0.73 & 1.82 \\
& & & & & & & & & & & & & & \\
{\sc new treatment of the energy equation} & & 0.05 & 20 & 0.605 & 7.20 & & 88 & 15 & 6 & 2 & & 0.21 & 0.39 & 0.63 \\
& & 0.10 & 20 & 0.629 & 6.20 & & 67 & 7 & 9 & 4 & & 0.23 & 0.46 & 0.94 \\
& & 0.25 & 20 & 0.623 & 9.05 & & 98 & 12 & 13 & 5 & & 0.23 & 0.46 & 0.90 \\
& & & & & & & & & & & & & & \\\hline                  
\end{tabular}
\end{table*}

{\it The effect of varying the initial level of turbulence, when using a barotropic equation of state.} Table \ref{TAB:GLOBAL} records -- for each treatment of the thermodynamics and each initial level of turbulence -- the number of different realisations (${\cal N}_{_{\rm REAL}}$), the efficiency (i.e. the mean fraction of the core mass converted into stars after $300\,{\rm kyr}$, $\eta\equiv\sum\{M_{\!\star}\}/M_{_{\rm CORE}}$), and the mean number of stars formed from one core ($\overline{\cal N}_{\!\star}$). With a barotropic equation of state, increasing the initial level of turbulence (a) decreases the efficiency of star formation, $\eta$, and (b) increases the mean number of stars formed from a single core, $\overline{\cal N}_{\!\star}$. The efficiency is reduced by increased turbulence, because the outer diffuse parts of the core are more vigorously dispersed, and therefore after $300\,{\rm kyr}$ they have not yet had time to fall back into the core and be incorporated into stars. An increased initial level of turbulence increases the total number of stars formed from a single core, because it drives more vigorous local compression, and thereby creates more protostellar seeds (i.e. more lumps which are sufficiently dense to be gravitationally unstable). It follows that the mean stellar mass decreases with increasing turbulence.

{\it The effect of varying the initial level of turbulence, when using the new treatment of the energy equation.} With the new treatment of the energy equation, these monotonic trends disappear. The efficiency, $\eta$, and the mean number of stars formed, $\overline{\cal N}_{\!\star}$, are only weakly dependent on the initial level of turbulence.

{\it The effects of the new treatment of the energy equation.} On the other hand, switching from the barotropic equation of state to the new treatment of the energy equation reduces the efficiency, $\eta$, somewhat, and significantly increases the mean number of stars formed from a single core, $\overline{\cal N}_{\!\star}$. There are two physical effects at work here.

{\it Lower stellar masses.} First, the new treatment of the energy
equation promotes the condensation of very low-mass stars, by taking
proper account of the thermal history and environment of the gas. With
the new treatment of the energy equation, a small proto-fragment tends
to be cooler, and thereby more inclined to condense out. This is
because the new treatment takes account of the fact that, being of
lower mass (and probably also non-spherical), the column-density
inhibiting the cooling of a low-mass proto-fragment is lower; and
because it contracts more slowly, its heating rate is
lower. Consequently its temperature is lower -- at a given density --
than the one prescribed by the barotropic equation of state, since the
latter is based on the behaviour at the centre of a collapsing,
spherical, non-rotating $1\,{\rm M}_\odot$ protostar. This is illustrated in Fig. \ref{FIG:DISCS}, which shows exactly the same initial conditions, evolved first with the barotropic equation of state (top frames), and then with the new treatment of the energy equation (bottom frames). It is evident that disc fragmentation is far more advanced in the simulation using the new treatment of the energy equation.

{\it More ejections and lower efficiency.} Second, because in the simulations performed with the new treatment of the energy equation the number of stars formed is higher, but they are of lower mass, these stars are less effective at mopping up the residual gas (thereby increasing their mass), and more likely to be ejected by dynamical interactions with other stars. Hence the amount of mass converted into stars is somewhat reduced, and the efficiency, $\eta$, is lower.

\begin{figure*}
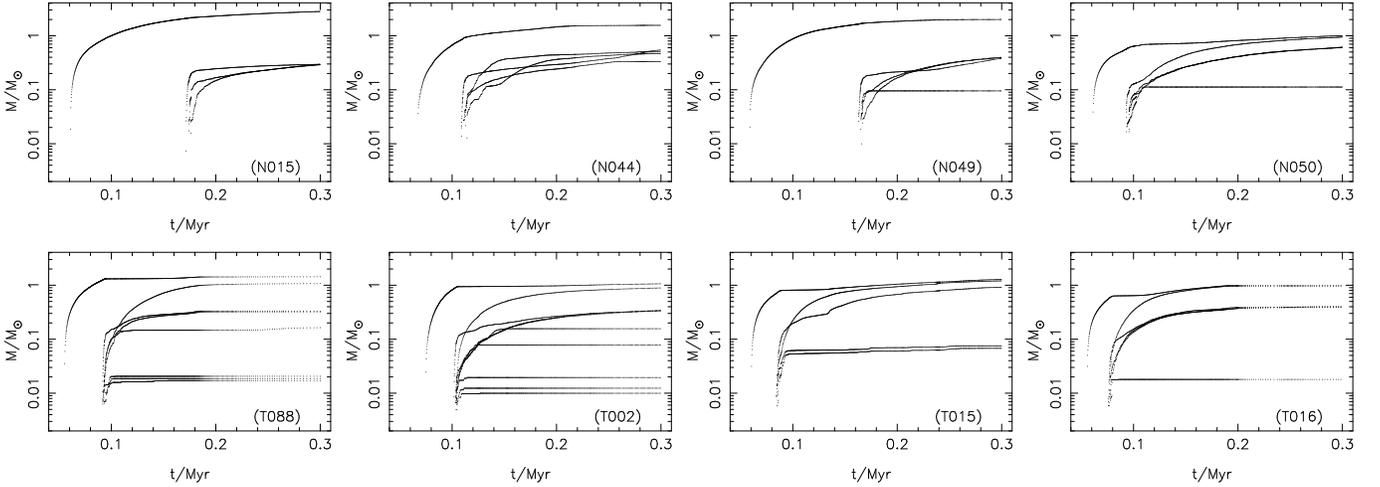

\centering$
\begin{array}{cccc}
\includegraphics[scale=0.2,angle=270]{N015nocol.ps} &
\includegraphics[scale=0.2,angle=270]{N044nocol.ps} &
\includegraphics[scale=0.2,angle=270]{N049nocol.ps} &
\includegraphics[scale=0.2,angle=270]{N050nocol.ps} \\
\includegraphics[scale=0.2,angle=270]{T088nocol.ps} &
\includegraphics[scale=0.2,angle=270]{T002nocol.ps} &
\includegraphics[scale=0.2,angle=270]{T015nocol.ps} &
\includegraphics[scale=0.2,angle=270]{T016nocol.ps} \\
\end{array}$
\caption{Stellar masses as a function of time, for a selection of simulations. Note (i) the delay between the formation of the primary and the formation of a clutch of secondaries (this is the time during which the circumprimary disc accumulates, until it becomes Toomre unstable); and (ii) the marked decline in the accretion rate onto the primary once the secondaries start to condense out.}
\label{FIG:ACCRETION}
\end{figure*}

{\it Disc fragmentation.} In fact, there is a common pattern of star formation in many of these simulations, irrespective of the treatment of thermodynamics. The low angular momentum material in the core collapses quickly to form the first star (hereafter the {\it primary}) on a timescale of $50\;{\rm to}\;70\,{\rm kyr}$, i.e. a bit longer than the initial freefall time at the centre of the core. Then material with higher angular momentum forms a circumstellar disc around the primary. This circumprimary disc grows in mass (the rate of infall onto the disc is greater than the rate at which mass accretes from the inner disc onto the primary) until the disc becomes Toomre unstable and fragments to form multiple secondaries. The delay between the formation of the primary and  fragmentation of the circumprimary disc is typically between 20 and $100\,{\rm kyr}$. During this time the disc is accumulating mass. Once the disc becomes Toomre unstable it normally fragments to produce between 3 and 5 stars, in the space of a few kyr.

{\it Accretion histories.} This pattern of fragmentation is illustrated on Fig. \ref{FIG:ACCRETION}, where, for a selection of simulations, we plot stellar masses as a function of time. On most of these plots, we see the primary forming, then a delay whilst the circumprimary disc builds up, and finally -- when the circumprimary disc becomes Toomre unstable -- the formation of a clutch of secondaries. Some of these secondaries are quickly ejected, and therefore end up as brown dwarfs, but others remain in the disc and accrete sufficient mass to become hydrogen-burning stars. Occasionally some even grow bigger than the primary.

\begin{figure*}
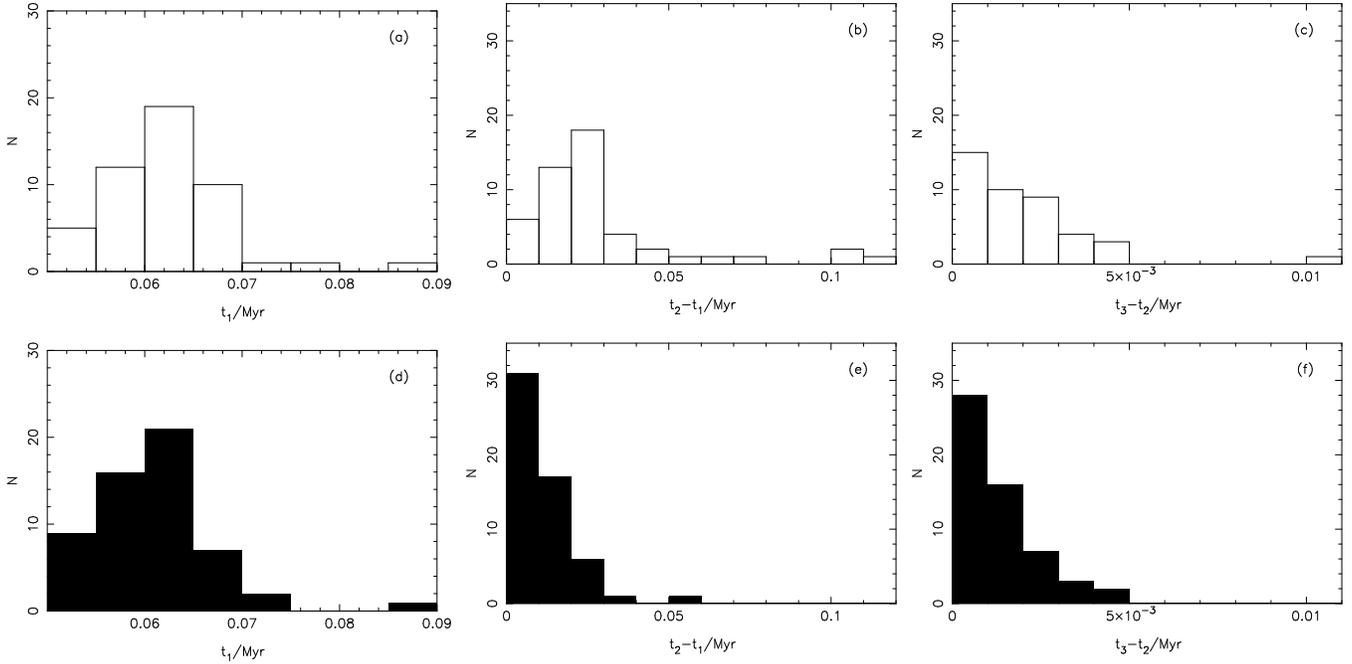

\centering$
\begin{array}{cccc}
\includegraphics[scale=0.25,angle=270]{dela.ps} &
\includegraphics[scale=0.25,angle=270]{delb.ps} &
\includegraphics[scale=0.25,angle=270]{delc.ps} \\
\includegraphics[scale=0.25,angle=270]{deld.ps} &
\includegraphics[scale=0.25,angle=270]{dele.ps} &
\includegraphics[scale=0.25,angle=270]{delf.ps} \\
\end{array}$
\caption{(a,d) The delay, $t_{_1}$, between the start of the simulation and the formation of the first star (the primary). (b,e) The delay, $t_{_2}-t_{_1}$, between the formation of the first and second stars. (c,f) The delay, $t_{_3}-t_{_2}$, between the formation of the second and third stars. The top row (a,b,c) is for the simulations performed using the barotropic equation of state, and the bottom row (d,e,f) is for the simulations performed using the new treatment of the energy equation.}
\label{FIG:DELAYS}
\end{figure*}

{\it The growth-time and fragmentation-time for the disc.} In Fig. \ref{FIG:DELAYS} we show the distributions of $t_{_1}$ (the time of formation of the first star); $t_{_2}-t_{_1}$ (the delay between the formation of the first and second stars); and $t_{_3}-t_{_2}$ (the delay between the formation of the second and third stars). $t_{_1}$ is the time it takes the low angular momentum material to assemble into the first stars and should be compared with the freefall time at the centre of the core ($\sim 40\,{\rm kyr}$); it has mean $\mu_{t_1}=61\,{\rm kyr}$ and standard deviation $\sigma_{t_1}=6.2\,{\rm kyr}$. $\;\;(t_{_2}-t_{_1})$ is the time it takes to assemble a Toomre-unstable disc around the primary; it has mean $\mu_{(t_2-t_1)}=20\,{\rm kyr}$ and standard deviation $\sigma_{(t_2-t_1)}=20\,{\rm kyr}$, so there is quite a large range in the times required to assemble a Toomre unstable disc. $\;\;(t_{_3}-t_{_2})$ is the time delay between the formation of the first two disc fragments; it has mean $\mu_{(t_3-t_2)}=1.6\,{\rm kyr}$ and standard deviation $\sigma_{(t_3-t_2)}=1.4\,{\rm kyr}$. This short mean $(t_3-t_2)$ reflects the fact that when the disc becomes unstable it becomes unstable over quite a large area, and therefore it tends to spawn several stars in quick succession.

{\it The effect of Toomre instability on the growth of the primary.} Another common feature of the accretion histories is that, when the circumprimary disc becomes Toomre unstable and fragments, the accretion rate onto the primary declines rapidly. Material which up until this juncture had been spiraling inwards and onto the primary star is now being used to create secondaries in the disc.

\begin{figure*}
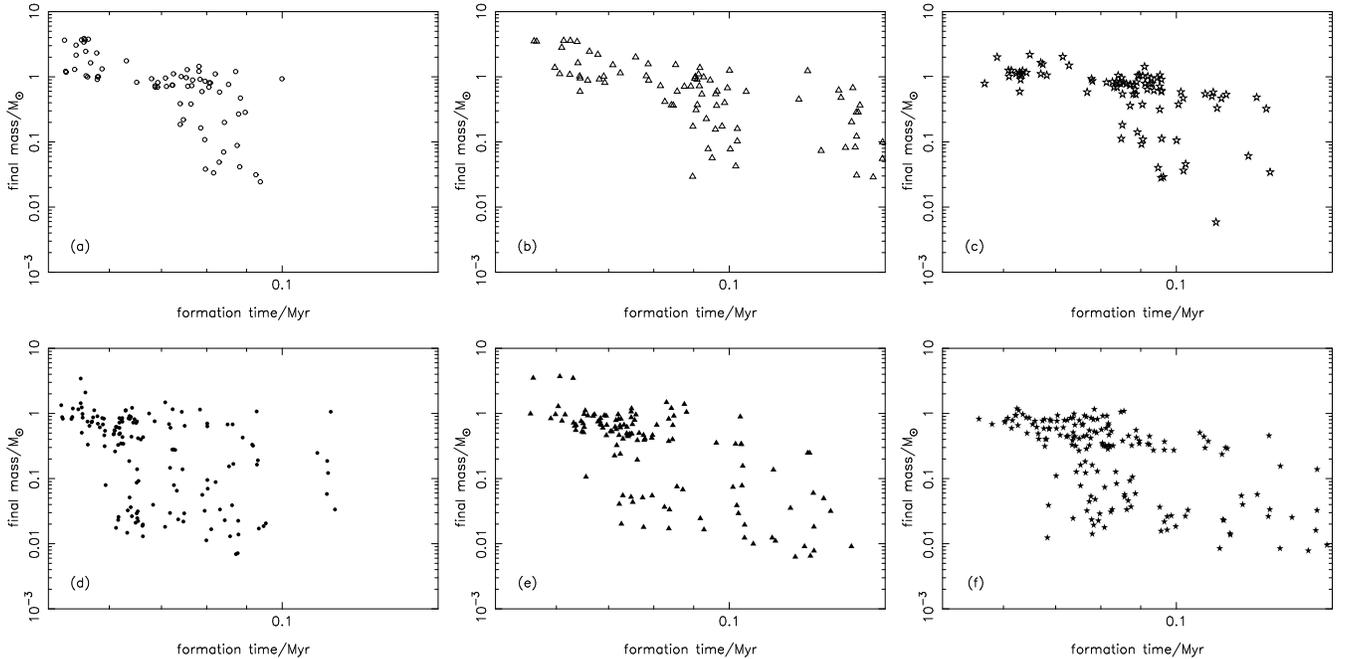

\centering$
\begin{array}{cccc}
\includegraphics[scale=0.25,angle=270]{tma.ps} &
\includegraphics[scale=0.25,angle=270]{tmb.ps} &
\includegraphics[scale=0.25,angle=270]{tmc.ps} \\
\includegraphics[scale=0.25,angle=270]{tmd.ps} &
\includegraphics[scale=0.25,angle=270]{tme.ps} &
\includegraphics[scale=0.25,angle=270]{tmf.ps} \\
\end{array}$
\caption{The final mass of each star (at $300\,{\rm kyr}$) against its formation time. The top row gives the results obtained with the barotropic equation of state for (a) $\alpha_{_{\rm TURB}}=0.05$, open circles; (b) $\alpha_{_{\rm TURB}}=0.10$, open triangles; and (c) $\alpha_{_{\rm TURB}}=0.25$, open stars. The lower row gives the results obtained with the new treatment of the energy equation for (d) $\alpha_{_{\rm TURB}}=0.05$, filled circles; (e) $\alpha_{_{\rm TURB}}=0.10$, filled triangles; and (f) $\alpha_{_{\rm TURB}}=0.25$, filled stars.}
\label{FIG:FINAL}
\end{figure*}

{\it Birth order.} Fig. \ref{FIG:FINAL} shows the final mass of every star plotted against its formation time. Although the more massive stars tend to form earlier, the correlation is fairly weak. In all cases there is a delay before any brown dwarfs form. This is because most of the brown dwarfs, and also some of the low-mass hydrogen-burning stars, form in discs around more massive stars, and these discs take time to accumulate.

\subsection{The mass distribution of protostars}

{\it Lower protostellar masses with the new treatment of the energy
  equation.} Material which is parked in a circumprimary disc has
  time to lose entropy -- to an extent that material which is
  compressed impulsively by turbulence does not. Consequently the
  masses of disc fragments are low, as predicted by Whitworth \&
  Stamatellos (2006), and demonstrated by detailed numerical
  simulations in Stamatellos et al. (2007b) and Stamatellos \& Whitworth
  (2008a,b). However, this effect can only be captured with the new treatment of the energy equation, since this treatment takes account of the slow rate of compressional heating for matter parked in the disc, and the relatively low local column-densities through which its cooling radiation has to diffuse. In contrast, the barotropic equation of state presumes that the matter is part of a spherical $1\,{\rm M}_\odot$ protostar, which by virtue of collapsing more rapidly is heated more vigorously, and has to cool through a larger column-density; therefore, at a given density, it is hotter and fragments less readily (i.e. into more massive fragments, if at all).

\begin{figure*}
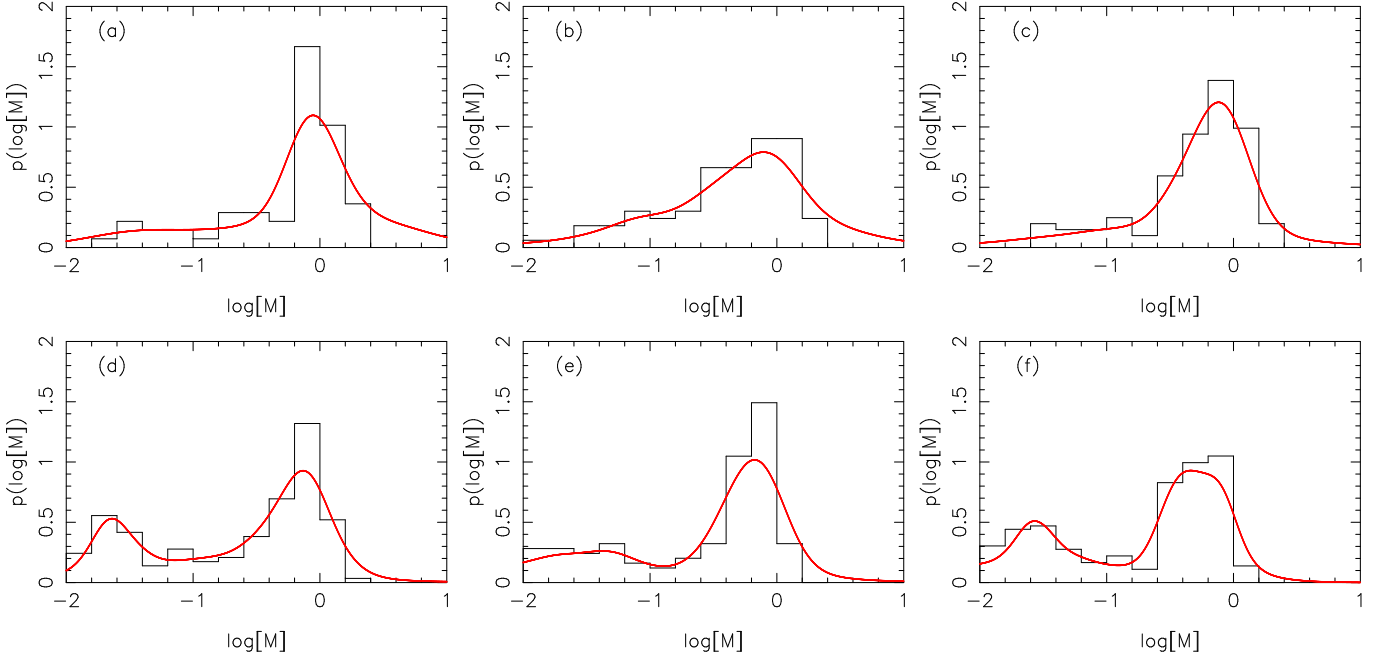

\centering$
\begin{array}{cccc}
\includegraphics[scale=0.27,angle=270]{mass1.ps} &
\includegraphics[scale=0.27,angle=270]{mass2.ps} &
\includegraphics[scale=0.27,angle=270]{mass3.ps} \\
\includegraphics[scale=0.27,angle=270]{mass4.ps} &
\includegraphics[scale=0.27,angle=270]{mass5.ps} &
\includegraphics[scale=0.27,angle=270]{mass6.ps} \\
\end{array}$
\caption{Normalised stellar mass distributions. The top row gives the mass distributions obtained with the barotropic equation of state for (a) $\alpha_{_{\rm TURB}}=0.05$, (b) $\alpha_{_{\rm TURB}}=0.10$, and (c) $\alpha_{_{\rm TURB}}=0.25$. The lower row gives the mass distributions obtained with the new treatment of the energy equation for (d) $\alpha_{_{\rm TURB}}=0.05$, (e) $\alpha_{_{\rm TURB}}=0.10$, and (f) $\alpha_{_{\rm TURB}}=0.25$. The black lines are histograms of the raw data, obtained using 15 equal logarithmic bins in the interval $\;-\,2\leq\log_{10}\left(M_{\!\star}/{\rm M}_\odot\right)\leq+\,1\,$, and the red lines are obtained by smoothing each protostellar mass with a Gaussian whose width is proportional to the separation between neighbouring masses (see Appendix A for details).}
\label{FIG:MASSES}
\end{figure*}

{\it Mass distributions.} The lower masses and greater numbers of stars formed with the new treatment of the energy equation predisposes the stars to mutual dynamical interactions which eject many of them before they have time to grow much by accretion. Fig. \ref{FIG:MASSES} shows the mass distributions obtained with the different combinations of thermodynamic treatment and initial level of turbulence. The black line shows the histogram obtained by distributing the final stellar masses into 15 logarithmic bins which are equally spaced in the interval
\[\hspace{0.3cm}-\,2\;\leq\;\log_{10}\!\left(\frac{M_{\!\star}}{{\rm M}_\odot}\right)\;\leq\;1\,,\hspace{1.8cm}\Delta\log_{10}\left(M_\star\right)\;=\;0.2\,.\]
The red line shows the mass distribution obtained when each stellar mass is smoothed using a Gaussian smoothing kernel with adaptive smoothing lengths dictated by the separation between masses (see Appendix A for details). Both the histogram, and the smoothed distribution, are normalised, in the sense that
\begin{eqnarray}
\int_{_{M=0}}^{^{M=\infty}}\,p_{_{\log_{10}(M)}}\,d\log_{10}(M)&=&1\,.
\end{eqnarray}
From Fig. \ref{FIG:MASSES} we see that the initial level of turbulence has little influence on the form of the mass distribution.

{\it Bimodal mass distributions.} However, switching from the barotropic equation of state to the new treatment of the energy equation not only increases the proportion of brown-dwarf stars formed, but actually produces a bimodal mass distribution. The larger mode comprises hydrogen-burning stars with masses concentrated in the range $0.3\;{\rm to}\;1.0\,{\rm M}_\odot$, whilst the smaller mode comprises brown dwarf stars with masses concentrated in the range $0.02\;{\rm to}\;0.06\,{\rm M}_\odot$. This smaller mode represents very low-mass stars formed by disc fragmentation (due to the enhanced cooling which low-mass fragments enjoy with the new treatment of the energy equation) and then ejected by mutual interactions (before they can grow much by accretion).

\subsection{Multiplicity}

{\it Measures of multiplicity.} To discuss the statistics of multiplicity, we adopt the conventions proposed by Reipurth \& Zinnecker (1993). We define {\it systems} to include single stars, and {\it multiple systems} to include only systems that contain more than one star. If $S$ is the number of single stars, $B$ the number of binary systems, $T$ the number of triple systems, $Q$ the number of quadruple systems, etc., then the total number of systems is $(S+B+T+Q+...)$, the total number of multiple systems is $(B+T+Q+...)$, and the total number of stars is $(S+2B+3T+4Q+...)$. We can then compute the multiplicity frequency, {\bf mf}, which measures the fraction of systems which are multiple, i.e.
\begin{eqnarray}
{\bf mf}&=&\frac{(B+T+Q+...)}{(S+B+T+Q+...)}\,;
\end{eqnarray}
the companion probability, {\bf cp}, which measures the fraction of stars which are in multiple systems, i.e.
\begin{eqnarray}
{\bf cp}&=&\frac{(2B+3T+4Q+...)}{(S+2B+3T+4Q+...)}\,;
\end{eqnarray}
and, from Goodwin et al. (2004b), the companion frequency, {\bf cf}, which measures the mean number of companions which a star has (irrespective of whether it is a primary), i.e.
\begin{eqnarray}
{\bf cf}&=&\frac{(2B+6T+12Q+...)}{(S+2B+3T+4Q+...)}\,.
\end{eqnarray}

{\it Multiplicity statistics.} In Table \ref{TAB:GLOBAL} we record --- for each ensemble of 20 simulations, representing a particular combination of thermodynamic treatment and initial level of turbulence --- the total numbers of singles ($S$), binaries ($B$), triples ($T$) and quadruples ($Q$) formed in all simulations; and the mean multiplicity frequency (${\bf mf}$), the mean companion probability (${\bf cp}$), and the mean companion frequency (${\bf cf}$).

\begin{figure}
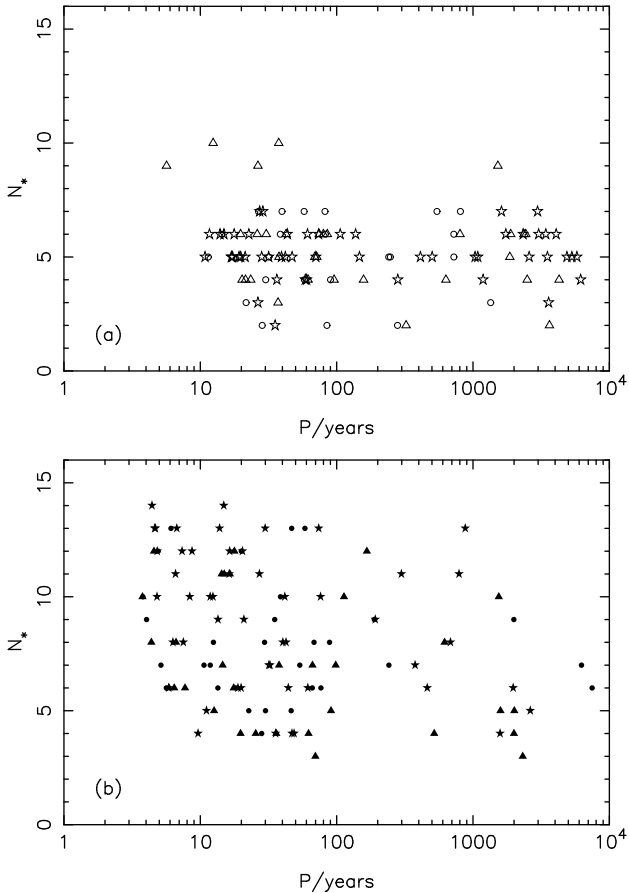

\centering$
\begin{array}{cccc}
\includegraphics[scale=0.35,angle=270]{NPNstar.ps} \\
\includegraphics[scale=0.35,angle=270]{TPNstar.ps} \\
\end{array}$
\caption{For each multiple system we plot the number of stars formed in that simulation, ${\cal N}_{\!\star}$, against the periods, $P$. (a) Results obtained using the barotropic equation of state; here open circles represent $\alpha_{\rm turb}=0.05$, open triangles represent $\alpha_{\rm turb}=0.10$, and open stars represent $\alpha_{\rm turb}=0.25$. (b) Results obtained using the new treatment of the energy equation; here filled circles represent $\alpha_{\rm turb}=0.05$, filled triangles represent $\alpha_{\rm turb}=0.10$, and filled stars represent $\alpha_{\rm turb}=0.25$.}
\label{FIG:PERIODS}
\end{figure}

{\it Computing periods.} Fig. \ref{FIG:PERIODS} shows, for each simulation, the number of stars formed in that simulation plotted against the periods of all the multiple systems identified at the end of the simulation. These periods are derived on the assumption that all multiple systems are hierarchical, which is not always true. Thus the two periods for a triple system are extracted by finding the period for the pair with the greatest specific binding energy, then treating this pair as a single star and finding the period of its orbit relative to the third star. This is appropriate for stable hierarchical systems, but of limited value for unstable non-hierarchical systems.

{\it Subsequent evolution of periods.} We should therefore expect some subsequent evolution in these distributions, with mutual interactions tending to lead to close systems becoming more tightly bound (occasionally with exchange of components) and wide systems being disolved. Eventually there will also be interactions with stars formed in neighbouring cores. These interactions will further disrupt the wider systems but have little effect on the closer systems. However, our simulations are not continued long enough for such interactions to be important.

{\it Period distributions.} The period distribution obtained using the barotropic equation of state has a mean $\mu_{\log_{10}(P/{\rm yr})}\simeq 2.2$ and a standard deviation $\sigma_{\log_{10}(P/{\rm yr})}\simeq 1.0$; periods range from $\,\sim\!10^4\,{\rm yr}$ down to $\,\sim\!10\,{\rm yr}$, and very few simulations form more than 7 stars. With the new treatment of the energy equation, the periods are on average shorter (by about a factor of 3), with a mean of $\mu_{\log_{10}(P/{\rm yr})}\simeq 1.7$, and a standard deviation $\sigma_{\log_{10}(P/{\rm yr})}\simeq 1.0$; periods range from $\,\sim\!10^4\,{\rm yr}$ down to $\,\sim\!3\,{\rm yr}$, and many simulations form more than 7 stars. 

{\it Effect of the new treatment of the energy equation on the period distribution.} The reason why the new treatment of the energy equation results in shorter-period multiples is that it allows the gas -- in particular, the gas in smaller proto-fragments -- to stay cooler to higher densities. Consequently the Jeans length, and hence the separations between neighbouring stars, tend to be smaller.

{\it Effect of the initial level of turbulence on the period distribution.} There is no obvious dependence of the period distribution on the level of turbulence, although this must be set against the poor statistics (between 26 and 53 periods for each combination of thermodynamics and initial level of turbulence).

{\it Unresolved orbits.} We should also caution that the low-period systems are poorly resolved, in the sense that at periastron the stars are closer together than $R_{_{\rm SINK}}$, and therefore their gravitational interaction is softenned. This means that they should probably be somewhat more tightly bound. We have checked the formation of the individual stars in some of these close systems, and established that in each case the two constituent stars (i.e. sinks) were initially created from well-defined and separate Jeans-unstable density peaks.

\begin{figure}
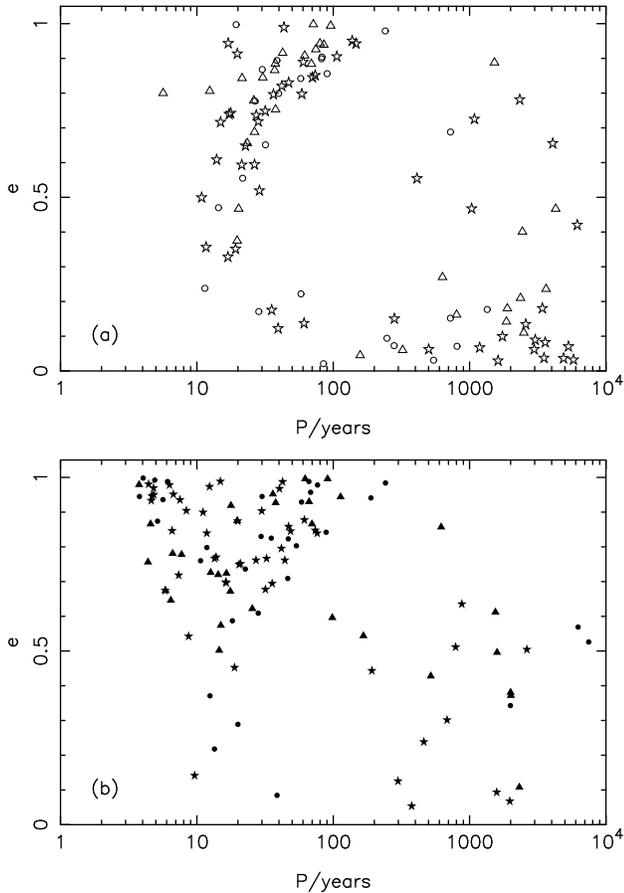

\centering$
\begin{array}{cccc}
\includegraphics[scale=0.35,angle=270]{NPe.ps} \\
\includegraphics[scale=0.35,angle=270]{TPe.ps} \\
\end{array}$
\caption{Orbital eccentricities, $e$, plotted against periods, $P$, for multiple protostars: (a) Results obtained using the barotropic equation of state; here open circles represent $\alpha_{\rm turb}=0.05$, open triangles represent $\alpha_{\rm turb}=0.10$, and open stars represent $\alpha_{\rm turb}=0.25$. (b) Results obtained using the new treatment of the energy equation; here filled circles represent $\alpha_{\rm turb}=0.05$, filled triangles represent $\alpha_{\rm turb}=0.10$, and filled stars represent $\alpha_{\rm turb}=0.25$.}
\label{FIG:ECCENTRICITIES}
\end{figure}

{\it Distribution of eccentricities.} Fig. \ref{FIG:ECCENTRICITIES} shows orbital eccentricities ($e$) plotted against periods ($P$), at the end of the simulations. The eccentricities are not strongly correlated with period, nor -- {\it modullo} the poor statistics (see above) -- do they appear to be correlated with the initial level of turbulence. However, there is a noticeable difference between the distributions obtained with the two different treatments of the thermodynamics. Using the barotropic equation of state, the distribution is concentrated towards high eccentricities, but there is still a substantial fraction, $\sim 25\%$, of systems having approximately circular orbits, $e\la 0.2$. Using the new treatment of the energy equation, the distribution of eccentricities is more strongly skewed towards high values, and less than 6\% have $e\la 0.2$.

{\it Effect of the treatment of thermodynamics on the distribution of eccentricities.} The barotropic equation of state facilitates the formation of low-eccentricity binaries by making it harder for circumbinary discs to fragment. At a given density the gas is hotter. Consequently, quite massive but relatively warm circumbinary discs resist further fragmentation and instead act to dampen orbital eccentricities by accreting slowly onto the existing binary components. In contrast, when the new treatment of the energy equation is used, massive circumbinary discs are relatively cool, so they fragment, and interactions between these additional fragments and the original components of the binary act to amplify the orbital eccentricities.

\begin{figure}
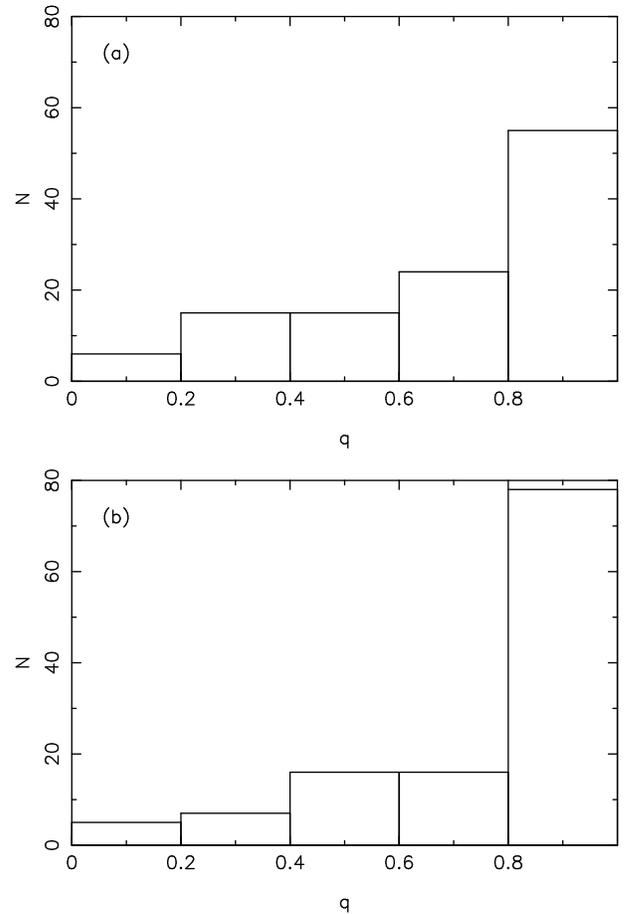

\centering$
\begin{array}{cccc}
\includegraphics[scale=0.35,angle=270]{qa.ps} \\
\includegraphics[scale=0.35,angle=270]{qb.ps} \\
\end{array}$
\caption{The distribution of mass ratios, $q$, for multiple protostars: (a) using the barotropic equation of state; (b) using the new treatment of the energy equation.}
\label{FIG:MASSRATIOS}
\end{figure}

{\it Distribution of mass ratios.} Fig. \ref{FIG:MASSRATIOS} shows the distributions of mass ratio, $q\equiv M_{_2}/M_{_1}$, at the end of the simulations. The distributions are strongly skewed towards $q\sim 1$, i.e. nearly equal component masses. The mass ratios do not appear to be correlated with the initial level of turbulence, $\alpha_{_{\rm TURB}}$, but again the statistics are poor. Mass ratios are correlated with orbital periods, in the sense that shorter-period systems tend to have higher mass-ratios, which is comparable with observations (e.g. Mazeh et al.1992). Since simulations conducted with the new treatment of the energy equation tend to produce multiples with shorter periods, they also tend to produce multiples with higher mass ratios.

{\it Mass equalisation in close binary systems.} A mechanism which drives mass ratios towards unity in simulations of star formation was first described by Chapman et al. (1992), and has subsequently been noted by Burkert \& Bodenheimer (1993) and by Bate \& Bonnell (1997) (but see Ochi et al. 2005 for a different view, and Clarke 2007 for a rebuttal of Ochi et al.). If a binary system continues to grow by accretion, the specific angular momentum of the infalling material (relative to the centre of mass of the binary system) tends to increase with time. Consequently the component with lower mass (the secondary, $M_{_2}$), which necessarily is on a more extended orbit, is better disposed to assimilate this material with high angular momentum, and therefore it grows in mass until it is comparable with the primary ($M_{_1}$). This is the mechanism that appears to be operating here. It is less effective in wide binary systems, because the components of a wide system tend to accrete from separate reservoirs.

{\it Competitive accretion and ejection.} We note that there is little evidence for competitive accretion in our simulations. The first star to form (the {\it primary}) is more often than not the most massive at the end. However, stars forming later in the simulation frequently acquire comparable, and occasionally greater, masses than the primary. The material which ends up in these stars is normally rather coherently located. For example, once the circumprimary disc has formed, the material destined to form a particular secondary star accumulates in a particular range of radii, and sits there until it is mopped up by the growing secondary star. Ejection does play a role in separating some stars from the reservoir of material they might otherwise have accreted, and thereby creating very low-mass stars. However, the material which accretes onto a star was in general present at the star's inception; its self-gravity contributed to the condensation which triggered the formation of a sink by pushing the density above $\rho_{_{\rm SINK}}$.

\subsection{Missing physics}

The switch from the standard barotropic equation of state to our new more realistic treatment of the energy equation produces significant changes in the statistical properties of the stars resulting from the collapse and fragmentation of an isolated, low-turbulence, $5.4\,{\rm M}_\odot$ core. However, there are several important physical effects missing from our simulations. In particular, there is no feedback from the stars, there are no (non-ideal) MHD effects, and the use of sink particles raises some concerns.

{\it Feedback.} Feedback from stars can take several forms.

(i) The radiation from the stars will heat the surrounding dust and
gas. Krumholz, Klein \& McKee (2007) have recently simulated the
  collapse and fragmentation of more massive cores ($100$ and
  $200\,{\rm M}_\odot$) with much higher initial levels of turbulence
  than those invoked in our simulations. Their treatment of the
  thermodynamics takes account not only of the energy equation and the
  transport of cooling radiation, but also of radiative feedback from
  the forming stars in the core. They find that their cores only spawn
  a small number of stars. This is because the primary protostar,
  which forms early on from material with relatively low angular
  momentum, has a high luminosity, and therefore stabilises the inner
  parts of its circumstellar accretion disc, by heating them
  up. Fragmentation is only possible in the outer more diffuse parts
  of the disc. We expect a similar -- but more modest -- effect in
  low-mass, low-turbulence cores, more modest because the primary
  luminosity will be much smaller. Nonetheless, it is likely that,
  even in the low-mass regime, the luminosity of the primary star is
  sufficient to inhibit fragmentation of the inner disc. The analytic
  work of Whitworth \& Stamatellos (2006) predicts that a disc around
  a Sun-like primary is unlikely to fragment inside $\sim\! 100\,{\rm
    AU}$, and this is confirmed by the simulations of Stamatellos et
  al. (2007b) and Stamatellos \& Whitworth (2008a,b). Consequently, the primary will end up more massive (by accreting the matter which is unable to fragment); the circumprimary disc will take longer to grow to Toomre instability; and the secondaries which then condense out of it will be smaller in number, and at larger radii.

(ii) Mechanical feedback, in the form of bipolar outflows will punch holes in the core. Preliminary exploration of this phenomenon (Stamatellos et al. 2005) suggests that it does not greatly change the efficiency of star formation, but it does slow it down (i.e. the delay between the formation of the primary and the formation of the secondaries is  longer). This needs to be explored further.

(iii) Ionising radiation and winds from massive stars produce more violent feedback. We have recently developed the numerical machinery to explore this (Bisbas et al., in preparation), but it is not part of the star-formation mode with which we are concerned here.

{\it MHD.} Non-ideal MHD effects are likely to play an important role, and we have developed a code to treat them (Hosking \& Whitworth 2004). However, it is a rather crude and inefficient code, and further work is ongoing to improve it to the stage where it can be used to perform a large ensemble of simulations. Price \& Bate (2008) have simulated the collapse and fragmentation of a massive magnetised turbulent core, using an ideal MHD code, with a barotropic equation of state. They find that the magnetic field reduces both the efficiency of star formation (i.e. the fraction of the core mass which ends up in stars) and the production of brown dwarfs. In an earlier paper (Price \& Bate 2007), using more idealised initial conditions (a spherical uniform-density cloud with an imposed $m=2$ perturbation), they have shown that a magnetic field can also inhibit disc fragmentation, by slowing the rate of disc growth and accelerating the rate at which angular momentum is redistributed.

{\it Sinks.} Finally, we note that the use of sinks may compromise our results, in ways which are hard to quantify (e.g. Commer\c{c}on et al. 2008). First, the use of sinks means that all processes on scales below $\sim\! 2R_{_{\rm SINK}}\!=\!10\,{\rm AU}$, and at densities above $\rho_{_{\rm SINK}}=10^{-11}\,{\rm g}\,{\rm cm}^{-3}$, are at best not properly resolved (e.g. orbits), and at worst excised completely (e.g. the second collapse when molecular hydrogen dissociates). Second, the creation of sinks promotes $N$-body interactions, and hence ejections of stars, whilst suppressing dissipative interactions between, and mergers of, stars. One can postpone the creation of sinks until very high densities are reached. For example, Stamatellos et al. (2007b) use $\rho_{_{\rm SINK}}=10^{-2}\,{\rm g}\,{\rm cm}^{-3}$. However, this is very expensive computationally. This is an area in which a more sophisticated algorithm is needed.

\subsection{Comparison with observation}

Since we only treat a single core mass, with a single initial radius and a single initial density profile, and since -- as discussed in the preceding section -- there are several potentially critical physical effects which are not included in our simulations, we do not expect to reproduce all the observed statistical properties of real stars. Nonetheless, it is appropriate to rehearse the various counts on which the properties of stars formed in our simulations conform to, or diverge from, reality; and to speculate on the reasons.

{\it Mean number of stars per core, $\overline{\cal N}_\star$.}
  Our simulations form too many stars per core, and furthermore this
  over-production of stars is significantly exacerbated by the switch
  from the barotropic equation of state to the new, more realistic
  treatment of the energy equation. This is because the new treatment
  allows circumstellar discs, and low-mass fragments thereof, to stay
  cool to higher densities than the barotropic equation of state
  (which treats all gas as if it were at the centre of a collapsing,
  non-rotating $\,1\,{\rm M}_\odot\,$ protostar). The analytic results
  of Rafikov (2005), Matzner \& Levin (2005), Kratter \& Matzner
  (2006) and Whitworth \& Stamatellos (2006), and the numerical
  simulations of Krumholz, Klein and McKee (2007), Stamatellos et
  al. (2007b) and Stamatellos \&
  Whitworth (2008a,b) all suggest that the inclusion of radiative feedback reduces the number of stars formed, essentially by heating the inner disc, and thus suppressing Toomre instability by increasing the cooling time (Toomre 1964; Gammie 2001). Whitworth \& Stamatellos (2006) show that, given a solar-mass primary star at the centre of the disc, it can only fragment at large radii, $R\ga 100\,{\rm AU}$. The inclusion of mechanical feedback (Stamatellos et al. 2005) and/or a magnetic field (Price \& Bate 2008) is also likely to reduce the number of stars formed, and in particular the number of brown dwarfs, by reducing the rate of accretion onto the primary and its circumprimary disc. Indeed, these effects are probably essential to reduce the efficiency of star formation in low-mass cores to the levels infered from observation. These levels are typically $\,\sim\! 30\,\%\,$ (e.g. Nutter \& Ward-Thompson 2007; Simpson et al. 2008).

{\it Mass distribution.} The overall mass distribution produced
  by a single core, as a fraction of the core's total mass, is not
  constrained by observation; if it were, we would know how to map the
  observed core mass function into the stellar initial mass
  function. One interesting feature of our results is the suggestion
  that, amongst the stars spawned by a single core, there might be a
  bimodal distribution of masses, comprising primary stars formed
  relatively early on, and secondary stars of much lower mass formed
  somewhat later by disc fragmentation.
We also note that the mass of core that we are simulating ($\sim 5.4
M_\odot$) is rather larger than the average isolated core (e.g. Alv\'es et
al. 2007; Nutter \& Ward-Thompson 2007).  This further complicates any
attempts to map the results of these simulations onto the observed
distributions which are the sum of a variety of core masses, mostly somewhat
smaller than our core.

{\it Multiplicity.} The multiplicity frequency of the stars formed in our simulations, ${\bf mf}\sim 0.2$, is too low, especially for the higher-mass stars; for the brown dwarfs and very low-mass hydrogen-burning stars (those with $M_\star<0.1\,{\rm M}_\odot$) $\;{\bf mf}\sim 0.2\,$ is actually in the middle of the range inferred from the limited observations available (Burgasser et al. 2007; Luhman et al. 2007; Joergens (2008). The multiplicity frequency is expected to rise if the inclusion of extra physics reduces the number of stars formed from a single core. If this reduction is attributable to the suppression of fragmentation in the inner parts of circumprimary discs, then the simulations of Stamatellos et al. (2008b) suggest that it will increase the multiplicity frequency of the higher mass stars ($M_\star\!\sim\! {\rm M}_\odot$), and have little effect on the multiplicity frequency of the very low-mass stars ($M_\star\!<\!0.1\,{\rm M}_\odot$); the simulations would then accord better with the observed distribution of multiplicity frequency, which appears to be a monotonically decreasing function of primary mass (e.g. Joergens 2008).

{\it Binary periods.} The periods, $P$, of the binary systems formed in our simulations fall in the range $\,3\la \left[P/{\rm yr}\right]\la 10^4\,$. Systems with shorter periods cannot be resolved, because the gravitational fields of sink particles are softened at distances closer than $R\!=\!5\,{\rm AU}$. Systems with longer periods must either form in more extended cores than the one we have modelled here, or they must result from interactions between stars formed in separate cores. An encouraging feature of the multiple systems formed in our simulations is the fact that most of the very low-mass systems ($M_{_1}<0.1\,{\rm M}_\odot$) have periods in the range $10$ to $100\,{\rm yr}$, in good agreement with the separations of observed very low-mass systems (e.g. Joergens 2008).

{\it Mass ratios and eccentricities.} The mass ratios of the
  multiple systems formed in our simulations are concentrated towards
  high values. This again accords with what is observed for very
  low-mass systems (e.g. Burgasser et al. 2007; their Fig. 5a), but
  contrasts with the flatter distribution observed in higher-mass
  systems. The multiple systems formed in our simulations are also skewed towards much more
eccentric orbits than observed systems.  However, the eccentricity
distribution at birth is almost impossible to compare to the distribution in
older systems.  Firstly, close systems will be circularised by tidal and
other dissipative forces.  Secondly, wider systems will be subject to
encounters which will rapidly change the birth eccentricity distribution
beyond recognition (Parker et al., in preparation). Our simulations do not
  address these possibilities.

\subsection{Convergence}

We have repeated one of our simulations with 50,000 and 80,000 {\sc
  sph} particles, to check whether our simulations are converged, in a
statistical sense (i.e. whether the statistical distributions of
stellar parameters does not depend on the number of {\sc sph}
particles used). For this purpose we have chosen simulation T011,
which has an initial level of turbulence $\alpha_{_{\rm TURB}}=0.10$,
and uses the new treatment of the energy equation; the mean number of
stars formed with this combination is $\overline{\cal N}_\star=6.2$
(see Table \ref{TAB:GLOBAL}). In the original T011 simulation, with
just 25,000 {\sc sph} particles, 6 stars are formed. With 50,000 {\sc
  sph} particles 6 stars are again formed. With 80,000 {\sc sph}
particles 8 stars are formed. We stress that in this context
convergence can only be discussed in a statistical sense. This is
because, with the low initial levels of turbulence we are using, the
gravitational fragmentation that ensues is seeded from two
sources. There are the small density enhancements created by subsonic
converging flows due to the initial imposed turbulent velocity field;
these are reproducable when using different particle numbers. However,
there is also particle noise; this is not reproducable when using
different particle numbers. Therefore convergence can only be tested
fully by repeating the whole ensemble of simulations with higher
particle numbers, and this is not feasible with the computational
resources at our disposal. We are currently preparing a paper which
demonstrates convergence in a simulation of gravitational
fragmentation by using very carefully relaxed initial conditions; the
imposed perturbation (which is reproducable) is then able to dominate
particle noise in seeding gravitational fragmentation. These
simulations exhibit excellent convergence, as do the simulations of
Jeans instability presented by Hubber et al. (2006). We are therefore
confident that our code is capturing gravitational fragmentation
faithfully.

\section{Conclusions}

We have performed a large ensemble of SPH simulations of the collapse and fragmentation of an isolated, turbulent $5.4\,{\rm M}_\odot$ core, with a view to establishing how the statistical properties of the resulting stars are influenced by (i) different initial levels of turbulence, and (ii) different treatments of the thermodynamics. We consider three initial levels of turbulence, characterised by $\alpha_{_ {\rm TURB}}=0.05,\;0.10\;{\rm and}\;0.25$. We treat the thermodynamics firstly with a standard barotropic equation of state, and secondly with a new treatment of the energy equation which captures all the important energy modes of the gas and takes account of radiation transport and variations in the opacity.

Increasing the initial level of turbulence tends to reduce the efficiency of star formation, $\eta$ (i.e. the fraction of the core mass which is converted into stars after $300\,{\rm kyr}$), and to increase the number of stars formed by a single core, ${\cal N}_{\!\star}$, but the effect is very small, and all the other statistical properties of the stars formed are essentially independent of $\alpha_{_ {\rm TURB}}$.

A common pattern is observed in which the low--angular-momentum material in the core collapses to form the first star (the primary) after $50\;{\rm to}\;70\,{\rm kyr}$ (i.e. just a bit longer than the initial freefall time at the centre of the core), and then a massive disc builds up around the primary. As soon as this circumprimary disc becomes Toomre unstable (which may take from $20\;{\rm to}\;100\,{\rm kyr}$), it rapidly breaks up into a clutch of secondary stars. Those secondaries that are quickly ejected from the disc normally end up as brown dwarf stars, whereas those secondaries that stay in the disc -- and particularly those that stay in the inner disc -- tend to grow by accretion, and sometimes they even grow larger than the primary.

Switching from the standard barotropic equation of state to our new more realistic treatment of the energy equation has several systematic effects:
\begin{itemize}
\item{the efficiency of star formation ($\eta\equiv\sum\left\{M_{\!\star}\right\}/M_{_{\rm CORE}}$) is reduced significantly (by $\,\sim\!15\%$);}
\item{the number of protostars formed (${\cal N}_{\!\star}$) is greatly increased (by $\,\sim\!40\%$);}
\item{a higher proportion of brown dwarf stars is formed;}
\item{the mean period of multiple systems is reduced (by a factor $\sim 3$);}
\item{the orbital eccentricities of multiple systems tend to be higher;}
\item{the mass ratios of multiple systems tend to be higher (i.e. more nearly equal components).}
\end{itemize}

All these trends can be attributed to the fact that the barotropic equation of state assumes that all gas is at the centre of a collapsing spherical $1\,{\rm M}_\odot$ protostar, and therefore it becomes adiabatic at relatively low densities. In contrast, our new more realistic treatment of the energy equation allows the gas in low-mass proto-fragments to remain approximately isothermal to relatively high densities.

These simulations do not capture all the deterministic effects in star formation. In particular, we do not take account of radiative and mechanical feedback from stars, we do not include non-ideal magneto-hydrodynamics effects, and we invoke sink particles (which must prejudice interactions between stars, in favour of elastic dynamical ejections, and against dissipation and mergers). We plan to explore these additional effects in subsequent papers. In particular, we anticipate that by including feedback we can reduce the efficiency of star formation from the very high levels produced here ($\,\sim\!60\%$) to values more compatible with observation ($\,\lesssim\!30\%$; Alves et al. 2007; Nutter \& Ward-Thompson 2007; Goodwin et al. 2008).

\begin{acknowledgements}
REA gratefully acknowledges the receipt of an STFC studentship
(PPA/S/S/2004/03981); DS and APW gratefully acknowledge the support of
an STFC Rolling Grant (PP/E000967/1); DS, SPG and APW acknowledge the
support of a Marie Curie Research Training Network
(MRTN-CT2006-035890). The colour plots in Fig. 1 were produced
  using SPLASH (Price 2007). We thank the referee for a constructive report which enabled us to improve an earlier version of this paper.
\end{acknowledgements}

\appendix
\section{Smoothed mass distributions}

The smoothed mass distributions on Fig. \ref{FIG:MASSES} are given by a sum of Gaussians,
\begin{eqnarray}
p_{\mu}\,d\mu&=&\sum_{i=1}^{i=I}\,\left\{\frac{1}{(2\pi)^{1/2}\,
\sigma_i}\,\exp\left[\frac{-\,(\mu-\mu_i)^2}{2\,\sigma_i^2}\right]\right
\}\,\frac{d\mu}{I}\,,\\\label{EQN:SIGMA}
\sigma_i^2&=&\left(\frac{\mu_I-\mu_1}{I-1}\right)^2\,+\,\left(\frac{\mu_
{i+2}-\mu_{i-2}}{4}\right)^2\,,
\end{eqnarray}
where $\mu\equiv\log_{10}\left(M\right)$ and $\mu_i\equiv\log_{10}\left(M_i\right)$. The standard deviation $\sigma_i$ is evaluated by adding -- in quadrature -- the mean separation between all masses across the entire mass spectrum (this is the first term on the righthand side of Eqn. \ref{EQN:SIGMA}), and the mean separation between the five nearest masses (this is the second term on the righthand side of Eqn. \ref{EQN:SIGMA}). Thus $\sigma_i$ combines a global and a local contribution to the smoothing. This smoothing is essentially {\it ad hoc}, and is designed purely to enable us to extract the large-scale features of the mass distribution, which are otherwise lost in the rather noisy histograms. In particular we are interested in the apparent bi-modality in the mass distributions obtained with the new treatment of the energy equation.

\end{document}